\documentclass[apj]{emulateapj}
\usepackage{apjfonts}

\newcommand{\kunit}{\hbox{\ cm$^{2}$\,s$^{-1}$}}

\shorttitle{TURBULENT DIFFUSION AND DEUTERIUM CHEMISTRY}
\shortauthors{BELL ET AL.}

\begin{document}

\title{The Influence of Deuteration and Turbulent Diffusion on the Observed $\rm{D/H}$ Ratio}
\author{T.~A.~Bell,\altaffilmark{1} K.~Willacy,\altaffilmark{2} T.~G.~Phillips,\altaffilmark{1} M.~Allen,\altaffilmark{2} and D.~C.~Lis\altaffilmark{1}}
\altaffiltext{1}
{Department of Astronomy, California Institute of Technology, \\
 1200 E California Blvd, Pasadena, CA 91125; tab@caltech.edu\email{tab@caltech.edu}.}
\altaffiltext{2}
{Jet Propulsion Laboratory, California Institute of Technology, \\
 4800 Oak Grove Drive, Pasadena, CA 91109.}


\begin{abstract}
The influence of turbulent mixing on the chemistry of the interstellar medium has so far received little attention. Previous studies of this effect have suggested that it might play an important role in mixing the various phases of the interstellar medium. In this paper we examine the potential effects of turbulent diffusion on the deuterium chemistry within molecular clouds. We find that such mixing acts to reduce the efficiency of deuteration in these clouds by increasing the ionization fraction and reducing freeze-out of heavy molecules. This leads to lower abundances for many deuterated species. We also examine the influence of turbulent mixing on the transition from atomic hydrogen to H$_2$ and from atomic deuterium to HD near the cloud edge. We find that including turbulent diffusion in our models serves to push these transitions deeper into the cloud and helps maintain a higher atomic fraction throughout the cloud envelope. Based on these findings, we propose a new process to account for the significant scatter in the observed atomic D/H ratio for galactic sightlines extending beyond the Local Bubble. Although several mechanisms have been put forward to explain this scatter, they are unable to fully account for the range in D/H values. We suggest a scenario in which turbulent mixing of atomic and molecular gas at the edges of molecular clouds causes the observed atomic D/H ratio to vary by a factor of $\sim$2.
\end{abstract}



\section{Introduction}\label{Introduction}

Since the first detection of a deuterium-bearing molecule, DCN, by \citet{Jefferts1973}, it has become clear that the fraction of deuterium contained in molecules can be extremely high in the dense interstellar medium (ISM), with abundances of some deuterated species observed to be comparable to those of their hydrogenated counterparts. This is in stark contrast to the low atomic D/H ratio measured by absorption studies in the diffuse gas, which has an average value of $1.56\pm0.04$$\times$10$^{-5}$ (i.e., $15.6\pm0.4$ ppm) in the local ISM \citep{Wood2004}.

The discovery of multiply deuterated molecules provided remarkable evidence to support the high molecular D/H ratios observed. Based on the elemental D/H ratio, such species would be expected to have abundances of order 10$^{-10}$ or less, relative to their hydrogenated analogs. Yet doubly deuterated ammonia, ND$_2$H, was found to have an abundance ratio $\rm{[ND_2H]/[NH_3]}$ of 0.5\% in the dark cloud L134N \citep{Roueff2000} and 3\% in the prestellar core 16293E \citep{Loinard2001}. Similarly, $\rm{[D_2CO]/[H_2CO]}$ ratios ranging from 1--40\% have been observed in low-mass protostars and prestellar cores \citep{Ceccarelli1998, Loinard2002, Bacmann2003}. Subsequent detections of CHD$_2$OH \citep{Parise2002}, D$_2$S \citep{Vastel2003}, and D$_2$H$^+$ \citep{Vastel2004} showed similarly high abundance ratios. Amazingly, molecules bearing three deuterons have also been detected, the first of which, triply deuterated ammonia, was observed to have an abundance ratio $\rm{[ND_3]/[NH_3]}\sim0.1$\% in the dark cloud Barnard 1 \citep{Lis2002} and in the Class 0 protostar NGC\,1333 IRAS\,4A \citep{vanderTak2002}. Triply deuterated methanol, CD$_3$OH, has also been detected in the low-mass protostar IRAS 16293-2422 with an abundance ratio $\rm{[CD_3OH]/[CH_3OH]}\sim1$\% \citep{Parise2004}. Such extreme enhancement of the D/H ratio within molecules points to the dramatic efficiency of deuterium fractionation in cold dense molecular clouds, a process that was first proposed by \citet{Solomon1973}.
 
At the low temperatures found in the cores of molecular clouds ($\sim$10~K), the process of fractionation acts to minimize the zero-point vibrational energy of molecules by swapping constituent atoms for their heavier isotopes. Exchange reactions therefore lead to the preferential substitution of H by D in molecules. Most gas-phase chemistry proceeds via ion-molecule reactions, the most important ion of which is H$_3^+$. Deuteration of this species through the reaction
\begin{eqnarray}
 \rm{H_3^+ + HD} & \rightleftharpoons & \rm{H_2D^+ + H_2 + 232\ K} \nonumber
\end{eqnarray}
is also preferred at low temperatures \citep{Watson1976}. Reactions with abundant gas-phase species, such as CO, O, and N$_2$, transfer the deuteron to other molecules and limit the amount of H$_2$D$^+$ that can be maintained, e.g.,
\begin{eqnarray}
 \rm{H_2D^+ + CO} & \rightarrow & \rm{DCO^+ + H_2}. \nonumber
\end{eqnarray}

In order for deuteration to proceed further and allow the formation of multiply deuterated species, H$_2$D$^+$ must survive long enough for it to undergo subsequent deuteron exchange:
\begin{eqnarray}
 \rm{H_2D^+ + HD} & \rightleftharpoons & \rm{D_2H^+ + H_2 + 187\ K}, \nonumber \\
 \rm{D_2H^+ + HD} & \rightleftharpoons & \rm{D_3^+  + H_2 + 234\ K}. \nonumber
\end{eqnarray}
This is possible when the atoms and molecules that destroy it are removed from the gas phase by freeze-out onto grain surfaces \citep{Brown1989a, Roueff2000, Roberts2000b}. This occurs in the cold dense regions of the ISM and the resulting depletion of abundant species, such as CO, has been observed in many sources \citep[e.g.,][]{Willacy1998, Kramer1999, Caselli1999, Bergin2002, Bacmann2002, Tafalla2004b}. Direct evidence for the link between freeze-out and enhanced deuteration comes from the observed correlation between CO depletion and the $\rm{[D_2CO]/[H_2CO]}$ ratio \citep{Bacmann2003}.

Grain-surface deuteration occurs by hydrogen abstraction and deuterium addition to mantle species on dust grains \citep{Tielens1983}. This requires a reservoir of atomic deuterium to be maintained in the gas. This, again, is produced as a result of the H$_3^+$$\to$H$_2$D$^+$$\to$D$_2$H$^+$$\to$D$_3^+$ series of reactions, with atomic deuterium a product of the dissociative recombination of these ions. For a more detailed discussion of deuterium chemistry, see, e.g., \citet{Roberts2003, Roberts2004}, \citet{Phillips2003}, and \citet{Phillips2006}.

Studies of the atomic D/H ratio in the local ISM provide an important constraint for models of galactic chemical evolution and insight into the processing of primordial gas up to the present era. Values for the galactic D/H ratio are obtained by deriving column densities of \ion{D}{1} and \ion{H}{1} from observed Lyman series absorption lines in interstellar gas toward background stars. Since the first of these measurements were carried out, an unexpectedly large variation has emerged for sightlines extending beyond the Local Bubble, with values ranging from 5 to 22 ppm \citep[see, e.g.,][for a recent discussion]{Linsky2006}. This significant scatter is at odds with the more uniform values obtained for the primordial D/H ratio, as probed by quasar absorption line systems and inferred from CMB anisotropy data. These methods have yielded a fairly consistent value of $(\rm{D/H})_{\rm{prim}}=27.5^{+2.4}_{-1.9}$~ppm \citep[see, e.g.,][]{Cyburt2003}.

Other than Big Bang nucleosynthesis, no formation mechanism for deuterium is known to exist, and explanations for the varying D/H ratio therefore focus on ways to destroy or hide the deuterium. A corollary of this is that the highest observed values for the D/H ratio must be closest to the true value. The process of astration---the conversion of deuterium to $^3$He, $^4$He, and heavier elements via stellar nucleosynthesis---has long been proposed to explain the generally low values for the atomic D/H ratio measured in the Galaxy, with differences in the degree of processing of the ISM being suggested as an explanation for the observed scatter. An alternative theory has been proposed by \citet{Draine2004, Draine2006}, in which deuterium atoms deplete onto dust grains in quiescent regions of the interstellar medium and are subsequently returned to the gas phase if the grain mantles are destroyed. The observed correlation between metal depletion and decrease in the D/H ratio has been used to support this theory \citep{Prochaska2005, Linsky2006, Steigman2007}, which predicts that regions that have experienced some degree of dust destruction should display both higher gas-phase abundances of refractory metals (titanium, iron, silicon, etc.) and higher D/H ratios (owing to the removal of both metals and deuterium from the grains). There are, however, outliers in these trends that disrupt the correlation and cannot be explained by simple dust depletion models. One important process acts to \textit{increase} the D/H ratio, namely the galactic infall of nearly pristine (i.e., unprocessed) material \citep[see, e.g.,][]{Prodanovic2008, Tosi1996}. If the infalling material is not fully mixed with the existing ISM, these inhomogeneities can also lead to variations in the observed D/H ratio.

It is well established that molecular clouds are turbulent, and that turbulence plays an important---perhaps even dominant---role in governing the dynamics and support of these clouds. What is less certain is the extent to which turbulent eddies can transport the gas and dust within a cloud, disrupting the abundance profiles predicted by static chemical models and redistributing the atomic and molecular species. The effects of turbulence on cloud chemistry were first considered by \citet{Phillips1981}. Subsequent work on this subject \citep[e.g.,][]{Boland1982, Federman1991, Xie1995, Willacy2002, Lesaffre2007} has demonstrated that turbulent mixing under conditions that are typical of molecular clouds can lead to ``smearing out'' of the abundance gradients, reducing the variation with depth as the gas is transported between the cloud envelope and its interior.

Turbulent transport in molecular clouds is a diffusive process \citep[often called eddy diffusion; see, e.g.,][]{Hinze1975} and the model employed in this work is derived from atmospheric chemical models in which the rate of turbulent transport for a given molecule depends on its abundance gradient. Such models describe the turbulent mixing between adjacent zones as a Fickian diffusion process with a diffusion coefficient $K$ derived from mixing length theory \citep{Taylor1915, Prandtl1925}, a phenomenological approach analogous to the concept of the mean free path in thermodynamics. For full details of this model, the reader is referred to the description contained in the original paper by \citet{Xie1995}.

As already discussed, the process of deuterium fractionation is extremely efficient at altering the D/H ratio of molecules in the cold, dense cores of molecular clouds. Indeed, \citet{Phillips2006} suggest that this process may lead to a significant fraction of deuterium being locked in molecules in cloud centers. Meanwhile, at the edges of these clouds and in diffuse regions of the ISM, where the gas is warm and purely atomic, the observed atomic D/H ratio should reflect the ``true'' elemental value (neglecting, for the moment, the possibility that deuterium may be also depleted onto dust grains). Turbulent diffusion serves to mix these two phases of the ISM---the warm diffuse material at the edge of clouds and the cold dense material in their centers---and in so doing, may alter the atomic D/H ratio observed at the cloud edge, as well as that in the cloud center. If this is the case, then the observed atomic D/H ratio along a particular line of sight will depend on a number of factors: 1) the degree of astration that the material along that sightline has been subjected to; 2) the amount of deuterium that has been depleted onto dust grains; 3) the efficiency of chemical fractionation in the gas along the sightline; 4) the degree of turbulent mixing of the different phases of the ISM; 5) the proximity of the line of sight to a molecular cloud. In general, these mechanisms act to {\it lower} the value of the observed atomic D/H ratio. Therefore, the highest observed value must be closest to the true elemental abundance ratio.

We have examined the effects of turbulent diffusion on deuterium chemistry within molecular clouds and the resulting change in the abundances of key deuterated species. We also propose a process by which the variation in the measured atomic D/H ratio along diffuse lines of sight can be explained by turbulent mixing of the gas observed along these sightlines with denser material in neighboring regions. We describe the modifications made to the existing model for this work and the physical parameters of the cloud models considered in \S\ref{The-Model}. The results of these models are presented in \S\ref{Results} and their implications for the observed D/H ratio are discussed in \S\ref{D-to-H-Ratio}. Finally, we summarize our findings in \S\ref{Summary}.


\vspace{1mm}
\section{The Model}\label{The-Model}

\begin{deluxetable}{llll}
 \tablecaption{Initial Abundances\label{Tab:Initial-Abundances}}
 \tablehead{
  \colhead{Species} & \colhead{$n_i/n$} & \colhead{Species} & \colhead{$n_i/n$}}
 \startdata
  H      & 1.00                  & He     & 0.14 \\
  D      & 1.60$\times$10$^{-5}$ & C$^+$  & 7.30$\times$10$^{-5}$ \\
  N      & 2.14$\times$10$^{-5}$ & O      & 1.76$\times$10$^{-4}$ \\
  Mg$^+$ & 7.00$\times$10$^{-9}$ & Fe$^+$ & 3.00$\times$10$^{-9}$ \\[-2mm]
 \enddata
 \tablecomments{$n=n(\mathrm{H})+2n(\mathrm{H_2})$.}
\end{deluxetable}

\begin{deluxetable}{ccccc}
 \tablecaption{Cloud Parameters\label{Tab:Cloud-Parameters}}
 \tablehead{
  \colhead{}      & \colhead{$r_0$} & \colhead{$n_0$} & \colhead{$T_{\rm{out}}$} & \colhead{$T_{\rm{in}}$} \\
  \colhead{Model} & \colhead{(pc)}  & \colhead{(cm$^{-3}$)} & \colhead{(K)}          & \colhead{(K)}}
 \startdata
  \#1 & 1.11 & 888 & \phm{0}10 & 10 \\
  \#2 & 4.00 & 250 & \phm{}100 & 10 \\[-2mm]
 \enddata
\end{deluxetable}

The models used in this work are extensions of those described in \citet{Xie1995} and \citet{Willacy2002}. The reader is referred to those papers for a detailed description of the implementation of turbulent diffusion in the model.

The diffusive transport appears as an additional term in the continuity equation for each species, at each depth-step,
\begin{equation}
 \frac{\partial n_i}{\partial t} = P_i - L_i - \frac{\partial \phi_i}{\partial r},
\end{equation}
where $n_i$ is the number density of species $i$, $P_i$ and $L_i$ are its production and loss terms, respectively, and $\phi_i$ is its net transport flux in the radial direction, given by:
\begin{equation}
 \phi_i\ \hbox{(cm$^{-2}$\,s$^{-1}$)} = -Kn\frac{\partial x_i}{\partial r},
\end{equation}
where $n$ is the total proton number density [$n=n(\mathrm{H})+2n(\mathrm{H_2})$], $x_i=n_i/n$ is the fractional abundance of species $i$, and $K$ is the turbulent diffusion coefficient (cm$^{2}$\,s$^{-1}$). All quantities are depth- and time-dependent, with the exception of the diffusion coefficient, which we assume to be constant across the cloud.

The strength of turbulent mixing in the model is governed by the turbulent diffusion coefficient $K$. It is difficult to accurately determine values for $K$, since the nature of turbulence in molecular clouds is still unclear. However, it is possible to derive an order-of-magnitude estimate for $K$ by assuming $K \sim \langle V_{\rm{t}}L \rangle$, where $V_{\rm{t}}$ is the turbulent velocity and $L$ is the correlation length of the diffusive process \citep[see][for details]{Xie1995}. Based on estimates of the correlation length in molecular clouds (0.1--0.5~pc) and typical turbulent velocities (0.1--1~km\,s$^{-1}$), we consider values of $K$ between 0 (i.e., no diffusion) and $10^{23}\kunit$.

In this work, our aim is to explore the potential impact of turbulent mixing on deuterium fractionation, rather than to make quantitative predictions for the resulting changes in abundances or fractionation ratios. Precise values of $K$ are therefore not important and we focus instead on exploring the plausible range of values that might be expected.

The chemical network used in the model includes gas-phase and grain-surface reactions, freeze-out onto grain surfaces, and evaporation of mantle species. We adopt the gas-phase reaction rates of the \textsc{rate99} release of the UMIST database for astrochemistry \citep{LeTeuff2000}. The treatment of freeze-out, grain-surface chemistry, and desorption mechanisms are described in \citet{Willacy2007}. We include the effects of thermal evaporation and cosmic ray induced desorption, and consider grain-surface reactions involving the addition of atoms and CH, OH, and NH radicals to grain mantle species, as well as their deuterated analogs. The migration of H and D across grain surfaces is assumed to be governed by the relatively slow hopping between surface sites. Using slow rates with a rate equation model has been shown by \citet{Vasyunin2009} to give good agreement with more exact Monte Carlo models of grain-surface chemistry. The model also considers the effects of an external interstellar radiation field (ISRF) on the chemistry, including photodissociation and photoionization of gas-phase species and photoevaporation of mantle species from grain surfaces. Self-shielding by H$_2$ increases the attenuation of dissociating far-ultraviolet (FUV) radiation, causing its photodissociation rate to drop rapidly with increasing depth into the cloud. We adopt the standard ISRF determined by \citet{Draine1978} and the H$_2$ self-shielding treatment of \citet{Lee1996}. HD does not self-shield, but some of its lines do overlap with those of H$_2$. This means that H$_2$ can provide some shielding. Here we follow the approach of \citet{Barsuhn1977}, who estimated that the overlap of HD and H$_2$ lines would reduce the HD photodissociation rate by $1/3$, assuming that the overlapped HD lines are totally shielded by the H$_2$ lines.

The initial gas-phase elemental abundances used in all models are listed in Table~\ref{Tab:Initial-Abundances}. Initial abundances for all other species are assumed to be zero. We adopt the average elemental deuterium abundance determined for the Local Bubble \citep[1.6$\times$10$^{-5}$;][]{Wood2004}. Although this value is only valid locally, the true elemental deuterium abundance for the galactic disk is unknown (and likely to be higher). The effect of an increased deuterium abundance on the deuterium fractionation has been investigated by \citet{Roueff2007b}, who find that it leads to a corresponding linear increase in the deuteration of molecules.

\begin{figure}
 \plotone{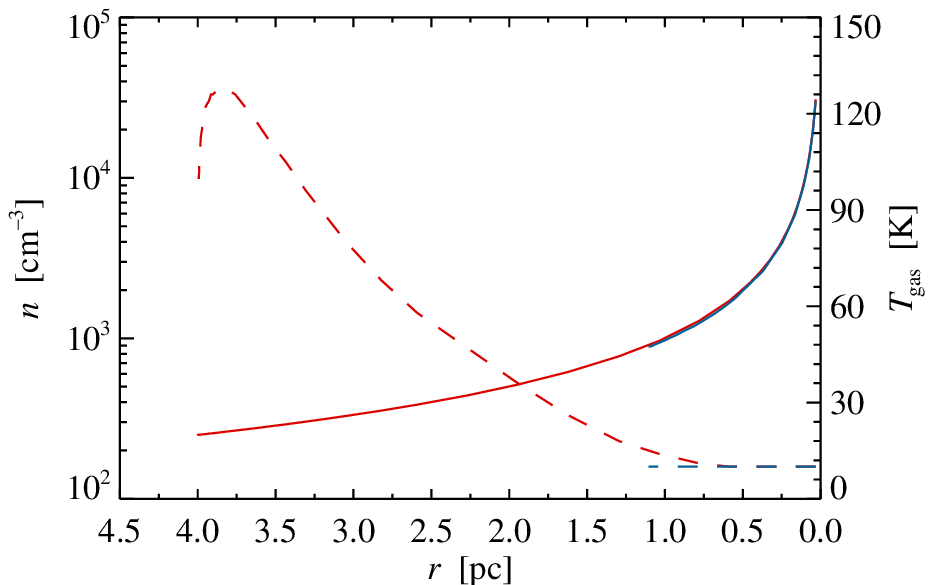}\\
 \plotone{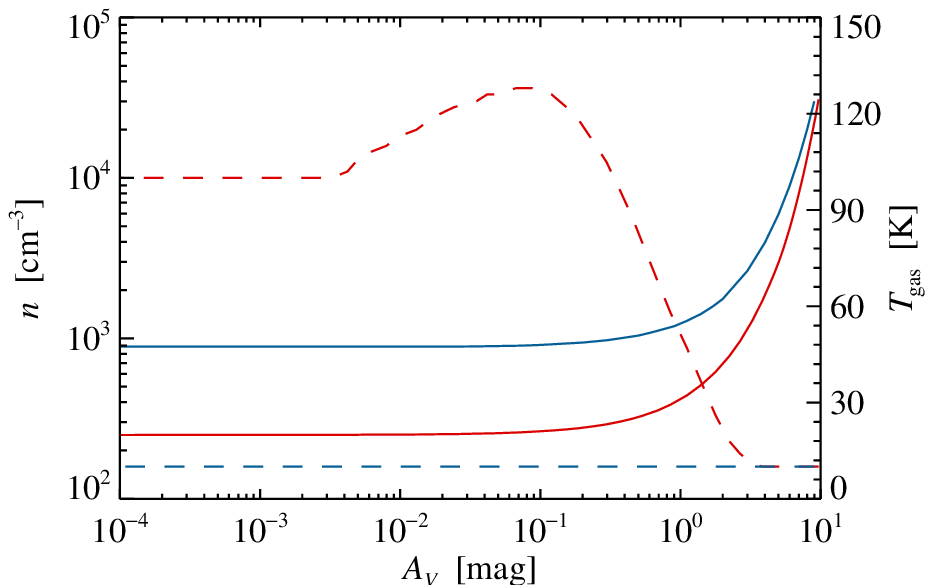}
 \caption{Cloud densities (solid lines) and gas temperatures (dashed lines) adopted in model \#1 (blue) and model \#2 (red). The top plot shows the radial dependence of these cloud properties and the bottom plot shows their variation as a function of visual extinction into the cloud.}
 \label{Fig:Cloud-Properties}
\end{figure}

\begin{figure*}
 \plottwo{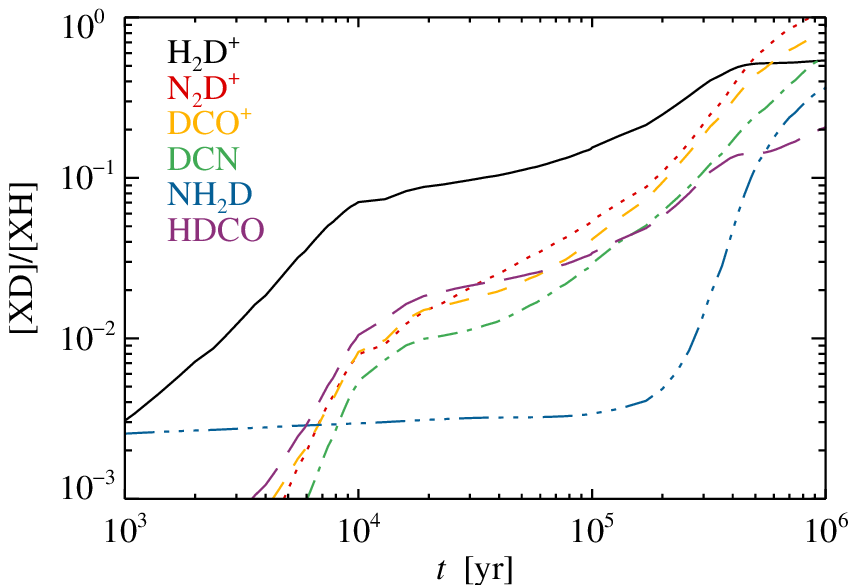}{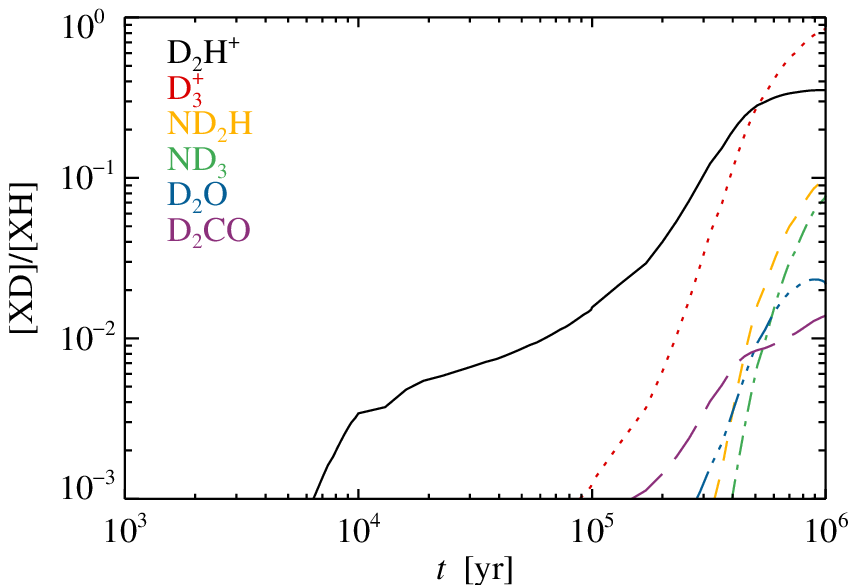}
 \caption{Fractionation ratios for various deuterated species at the cloud center, as a function of time. The left-hand plot shows results for a number of singly deuterated species, H$_2$D$^+$ (black, solid), N$_2$D$^+$ (red, dotted), DCO$^+$ (yellow, dashed), DCN (green, dot-dashed), NH$_2$D (blue, triple dot-dashed), and HDCO (purple, long-dashed). The right-hand plot shows results for various doubly and triply deuterated species, D$_2$H$^+$ (black, solid), D$_3^+$ (red, dotted), ND$_2$H (yellow, dashed), ND$_3$ (green, dot-dashed), D$_2$O (blue, triple dot-dashed), and D$_2$CO (purple, long-dashed). Results for model \#1 are shown.}
 \label{Fig:Central-Fractionation-Ratios}
\end{figure*}

\subsection{Changes to the Model}

The major modifications to the code necessary for this work were the introduction of deuterated species into the network of chemical reactions and the inclusion of freeze-out, grain surface chemistry, and desorption processes. The treatment of deuterium chemistry and these additional processes is described in detail in \citet{Willacy2007}.

\subsection{Cloud Properties}

In order to investigate the effects of turbulent diffusion under different interstellar environments, we have applied our code to two distinct sets of cloud parameters, which we list in Table~\ref{Tab:Cloud-Parameters}. We assume a density profile of the form $n(r)=n_0(r_0/r)$ in both models. Since the code does not attempt to solve the thermal balance within the cloud, we impose fixed temperature profiles on the gas and dust. The models are distinct in the following ways. In model \#1, which is identical to that used previously by \citet{Xie1995} and \citet{Willacy2002}, we adopt a cloud size of $r_0=1.11$~pc (3.41$\times$10$^{18}$~cm), a number density at the cloud edge of $n_{0}=888$~cm$^{-3}$, and a constant gas and dust temperature of 10~K (i.e., $T_{\rm{in}}=T_{\rm{out}}=10$~K). Model \#2 attempts to simulate a more extended cloud, with a larger, diffuse envelope extending to $r_{0}=4.0$~pc (1.23$\times$10$^{19}$~cm) and a density at its edge of $n_{0}=250$~cm$^{-3}$. To account for the higher temperatures present in diffuse cloud envelopes, the temperature profile used in model \#2 was determined using a separate code, designed to calculate the chemical and thermal structure of photon-dominated regions (PDRs) by treating the depth-dependent chemistry, heating and cooling simultaneously (see \citealt{Bell2006a} for details). The same chemical network and incident radiation field were used in the PDR code and the density profile for model \#2 was adopted. The resultant temperature profile initially rises sharply from $T_{\rm{out}}=100$~K at the cloud edge to $\sim$130~K at a radius of 3.8~pc, before gradually falling to $T_{\rm{in}}=10$~K at the cloud center. The gas density and temperature profiles for models \#1 and \#2 are shown in Fig.~\ref{Fig:Cloud-Properties}. In both models, the dust temperature is assumed to be constant at 10~K throughout the cloud. We note that, while this may only be truly valid for the dense central regions, the choice of dust temperature in the cloud envelope has less of an impact on the chemistry because freeze-out is less effective at the lower densities found there.

In all models, the chemical evolution of the cloud was computed for a period of 1~Myr. The results are discussed in the next section, where all abundances are quoted relative to the total proton number density, $n=n(\mathrm{H})+2n(\mathrm{H_2})$.


\section{Results}\label{Results}

Before examining the effects of turbulent diffusion on the deuterium chemistry, it is important to compare the results of our static chemical model to previous theoretical work and to observations. Chemical models with increasingly sophisticated treatment of the deuterium fractionation and freeze-out have been constructed over the years \citep[e.g.,][]{Roberts2000a, Roberts2000b, Rodgers2001, Roberts2003, Flower2004, Aikawa2005, Flower2006a, Flower2006b, Roueff2007b}. These are generally time-dependent, but often limited to a single spatial point with conditions appropriate for the centers of dark clouds, characterized by low temperature ($T\sim10$~K), high density ($n\ge10^4$~cm$^{-3}$), and high visual extinction ($A_V\sim10$~mag). Most of these models do not consider the more extended envelopes of molecular clouds, where temperatures are higher and densities lower, and the material is exposed to the ISRF \citep[however, see][]{LePetit2002}. For the purposes of comparison, we show in Fig.~\ref{Fig:Central-Fractionation-Ratios} the fractionation ratios of a number of deuterated species at the center of our model cloud, as a function of time. In this paper, we use the term ``fractionation ratio'', or $\rm{[XD]/[XH]}$, to mean the abundance ratio of a deuterated species with respect to its fully hydrogenated counterpart. It can be seen that fractionation ratios for many deuterated species approach values of $\sim$10\% or higher at the cloud center after 1~Myr of chemical evolution. Particularly impressive are N$_2$D$^+$, which reaches a higher abundance than N$_2$H$^+$, and ND$_3$, whose abundance is comparable to ND$_2$H. In contrast, the fractionation ratios at the outer edge of the cloud are $\sim$0.1\% or less for most deuterated species.

The majority of models described in the literature employ much higher number densities than are considered here (typically of order 10$^6$~cm$^{-3}$, whereas our central cloud density is 3$\times$10$^4$~cm$^{-3}$). This makes a direct comparison difficult. \citet{Roberts2003} performed detailed chemical modeling that included multiply deuterated species and freeze-out of gas-phase species onto grain mantles. They considered models with $T=10$~K and $n(\rm{H_2})=10^4$~cm$^{-3}$, conditions very similar to those in the center of our model cloud. In order to compare our chemical model to theirs, we ran a single depth-step model with the same physical conditions and initial abundances as used in their paper. Comparing the fractionation ratios listed in their Table~1 with the values from our single-point model at $t=1$~Myr, the two models agree within a factor of 2 for most deuterated species, the only exception being ND$_3$, which is an order of magnitude higher in our model. This disagreement is probably due to differences in the choice of branching ratios in the chemical networks. A detailed analysis of the differences between our model and those of others is beyond the scope of this paper, but despite these minor differences, we consider our model to be in good agreement with the results of \citet{Roberts2003}.

When comparing our results to observational data, it is more appropriate to consider fractionation ratios based on total column densities through the model cloud, since observed fractionation ratios are determined in the same way. We therefore list in Table~\ref{Tab:Fractionation-Ratios} the fractionation ratios for various species in the model determined from total column densities, $N(\rm{XD})/N(\rm{XH})$, alongside the typical range of observed values quoted in the literature. The general agreement is very good, with model fractionation ratios for most species falling within, or very close to, the observed range of values. The notable exceptions are HDO, the bulk of which is expected to be in the form of ice, and D$_2$CO, which is typically observed in warmer sources or regions of higher density \citep{Bacmann2003}. The model values generally fall at the lower end of the observed ranges, which is to be expected, since the majority of the observed sources have much higher central densities.

Our intention in this paper is to examine the general effects of introducing turbulent diffusion into models of deuterium chemistry in the ISM, focusing on overall trends, rather than making quantitative predictions for specific sources. With this in mind, we regard the results of our deuterium model to be in good agreement with previous models and observations. 

\begin{deluxetable}{lccl}
 \tablecaption{Model Fractionation Ratios for Selected Species and Typical Observed Values\label{Tab:Fractionation-Ratios}}
 \tablehead{
  & \colhead{Model\tablenotemark{a}} & \colhead{Observed} & \\
  & \colhead{$N(\rm{XD})/N(\rm{XH})$} & \colhead{$\rm{[XD]/[XH]}$} & \\
  \colhead{Species} & \colhead{\%} & \colhead{\%} & \colhead{References}}
 \startdata
  H$_2$D$^+$ & \phm{}28\phm{.0} & \nodata        & \nodata \\
  D$_2$H$^+$ & \phm{}12\phm{.0} & \nodata        & \nodata \\
  $\rm{D_2H^+\!\!/H_2D^+}$
             & \phm{}42\phm{.0} & \phm{0}50--100 & 1\tablenotemark{b} \\
  N$_2$D$^+$ & \phm{}19\phm{.0} & 10--30         & 2, 3 \\
  DCO$^+$    & \phm{}15\phm{.0} & \phm{0}4--18   & 4, 5 \\
  DCN        & \phm{0}9\phm{.0} & 4--6           & 2, 6 \\
  DNC        & \phm{0}4\phm{.0} & 2--9           & 2, 7 \\
  NH$_2$D    & \phm{}10\phm{.0} & 10--30         & 5, 8, 9 \\
  ND$_2$H    & \phm{0}2\phm{.0} & 0.5--4\phm{.0} & 9, 10, 11 \\
  ND$_3$     & \phm{0}0.9\phm{} & 0.02--0.55     & 9, 12, 13 \\
  HDO        & \phm{0}8\phm{.0} & 0.2--3\phm{.0} & 14 \\
  D$_2$O     & \phm{0}0.2\phm{} & \nodata        & \nodata \\
  HDCO       & \phm{0}6\phm{.0} & 3--7           & 2, 6 \\
  D$_2$CO    & \phm{0}0.2\phm{} & \phm{0}1--10   & 15, 16 \\[-2mm]
 \enddata
 \tablenotetext{a}{Model fractionation ratios are determined from total column densities integrated over all radial points in model \#1 at 1 Myr.}
 \tablenotetext{b}{Observed fractionation ratios for H$_2$D$^+$ and D$_2$H$^+$ do not exist in the literature, but the $\rm{[D_2H^+]/[H_2D^+]}$ ratio has been measured in the prestellar core 16293E by \citet{Vastel2004}.}
 \tablerefs{(1) \citealt{Vastel2004}; (2) \citealt{Turner2001}; (3) \citealt{Gerin2001}; (4) \citealt{Williams1998}; (5) \citealt{Tine2000}; (6) \citealt{Roberts2002}; (7) \citealt{Hirota2003}; (8) \citealt{Hatchell2003}; (9) \citealt{Roueff2005}; (10) \citealt{Roueff2000}; (11) \citealt{Loinard2001}; (12) \citealt{Lis2002}; (13) \citealt{vanderTak2002}; (14) \citealt{Parise2005}; (15) \citealt{Bacmann2003}; (16) \citealt{Roberts2007}.}
\end{deluxetable}

\subsection{General Trends}\label{General-Trends}

We find that turbulent mixing can often lead to unintuitive effects on the deuterium chemistry, reducing the level of deuteration in some species, whilst enhancing that of others at certain points within the cloud. This is due to the complex interplay between mixing and chemical processes, redistributing important feed molecules that govern the deuterium chemistry, whilst disrupting the freeze-out of abundant species such as CO that can hinder deuteration of gas-phase species.

\begin{figure*}
 \plottwo{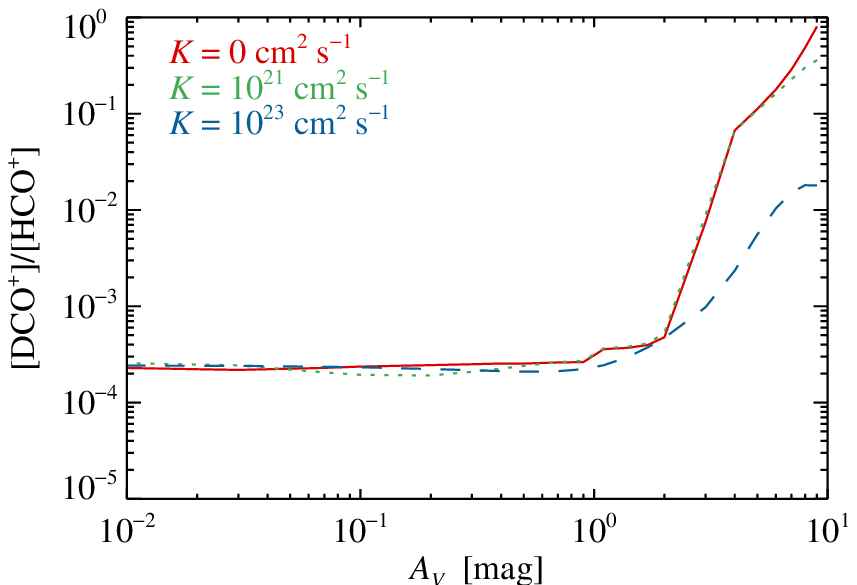}{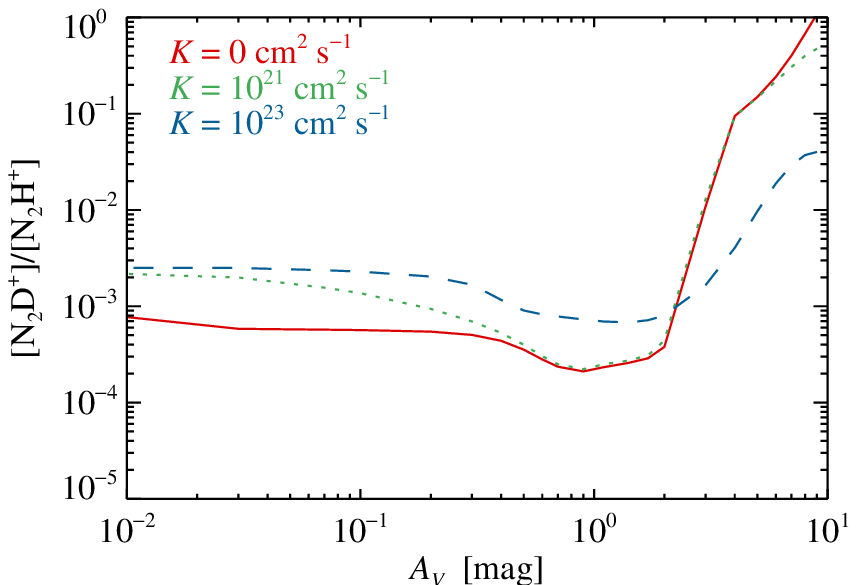}\\
 \plottwo{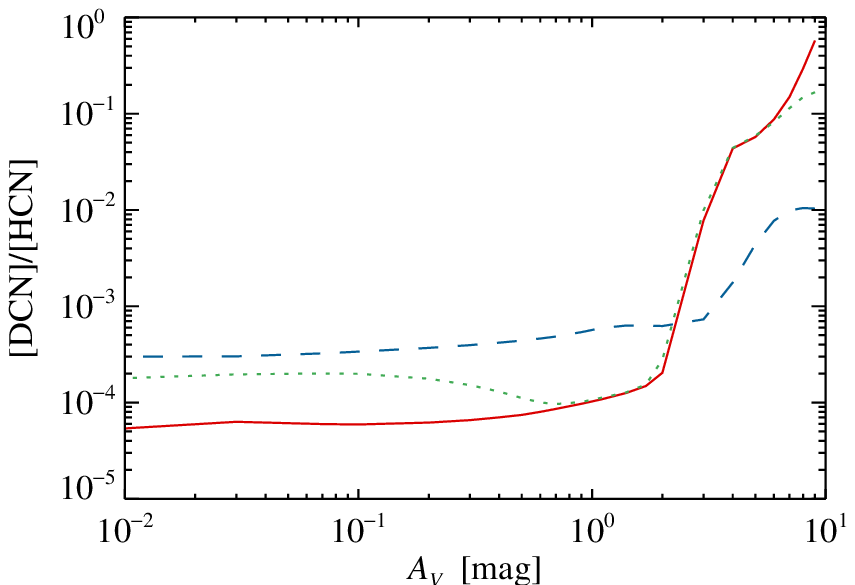}{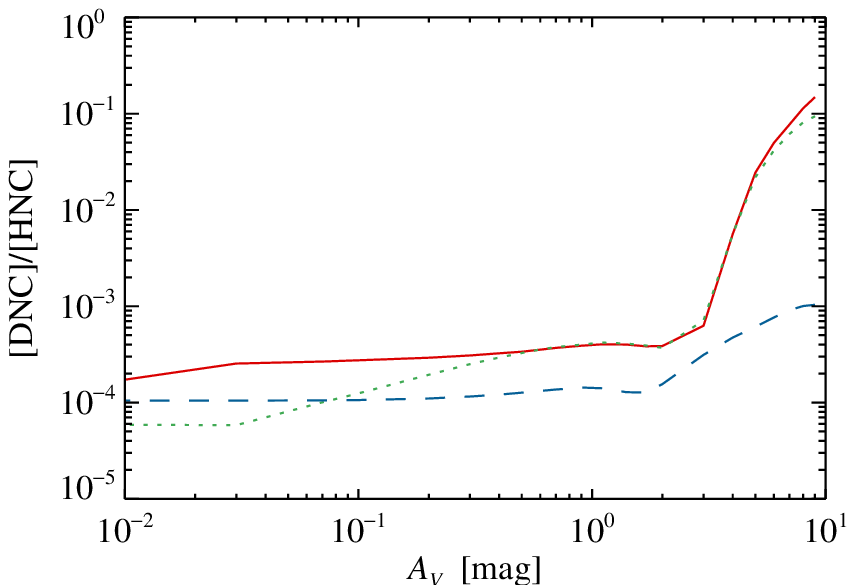}\\
 \plottwo{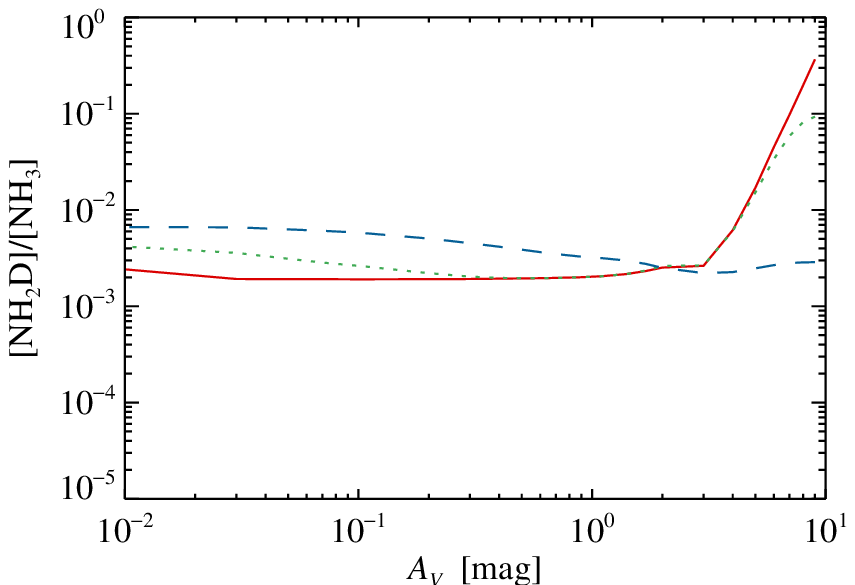}{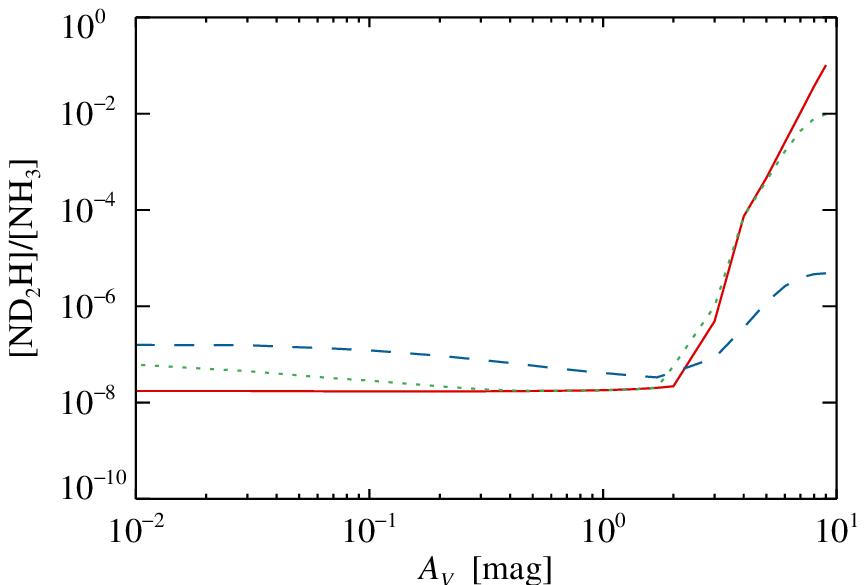}\\
 \plottwo{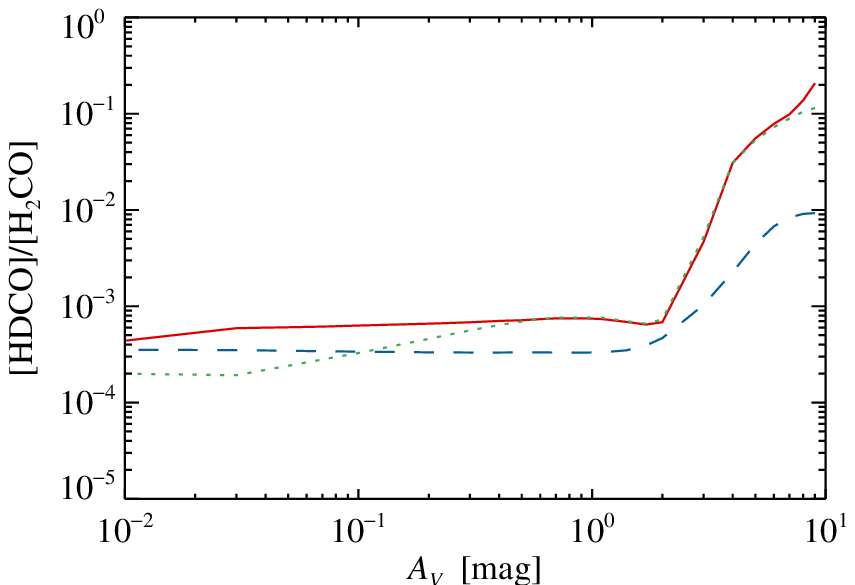}{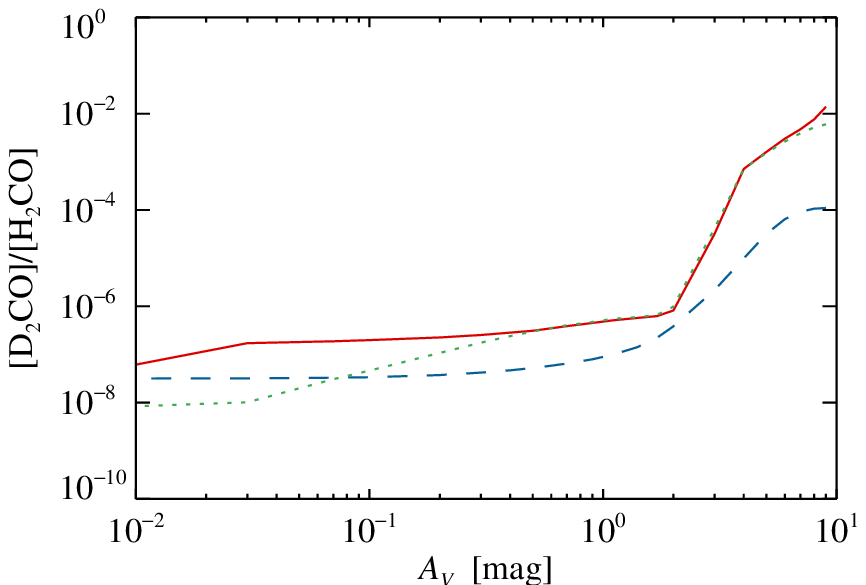}
 \caption{Fractionation ratios for DCO$^+$, N$_2$D$^+$, DCN, DNC, NH$_2$D, ND$_2$H, HDCO, and D$_2$CO (from top-left to bottom-right, respectively) for diffusion coefficients of $K=0$, $10^{21}$, and $10^{23}\kunit$ (red solid, green dotted, and blue dashed lines, respectively). All plots show results for model \#1 at $t=1$ Myr.}
 \label{Fig:Fractionation-Ratios}
\end{figure*}

To illustrate the predicted effects of turbulent diffusion on the deuterium chemistry, we show in Fig.~\ref{Fig:Fractionation-Ratios} the fractionation ratios of a number of important deuterated molecules (at $t=1$~Myr) as a function of visual extinction into the cloud for a set of models with diffusion coefficients of $K=0$ (i.e., no diffusion) to $K=10^{23}\kunit$ (an upper limit to the possible range of values). It is immediately obvious that increasing levels of turbulent diffusion generally act to reduce the abundances of deuterated molecules. In the cloud center ($A_V > 1$~mag), all deuterated species suffer under the influence of turbulent diffusion, with fractionation ratios dropping by up to several orders of magnitude. However, the effect is not uniform throughout the cloud, and in the envelope, the deuteration of some species actually increases when turbulent mixing is active. In the next two sections, we consider the effects of turbulent diffusion on the cloud center and envelope separately.

\subsection{Cloud Core}\label{Cloud-Core}

Turbulent diffusion limits the abundance of deuterated species in the cloud center by disrupting two of the key features that are important for effective deuteration:
\begin{enumerate}
\item The ionization fraction (e$^-$ abundance) is enhanced.
\item Depletion of gas-phase species onto grains is reduced.
\end{enumerate}
Together, these lead to significantly lower abundances of H$_2$D$^+$, D$_2$H$^+$, and D$_3^+$ compared to chemical models that do not include turbulent mixing, in turn resulting in lower abundances of the deuterated molecules that are formed from these ions.

\begin{figure}
 \plotone{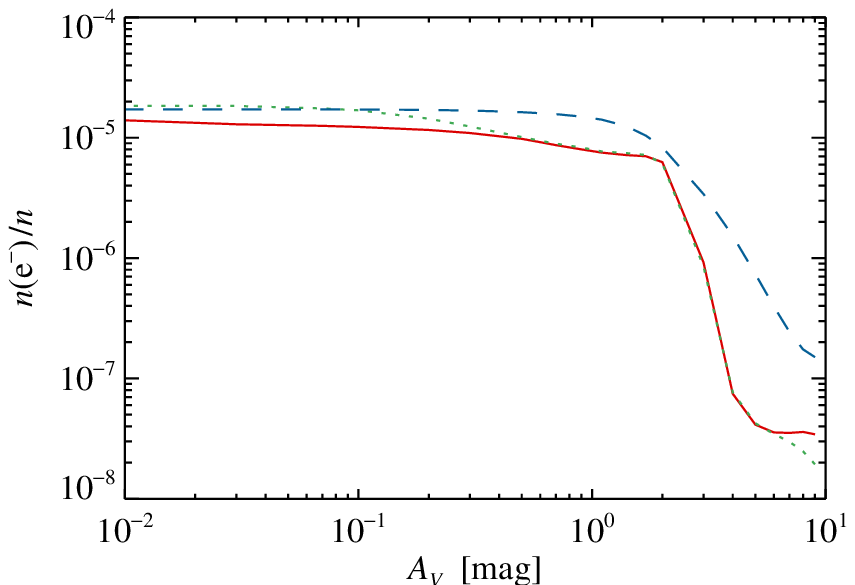}\\
 \plotone{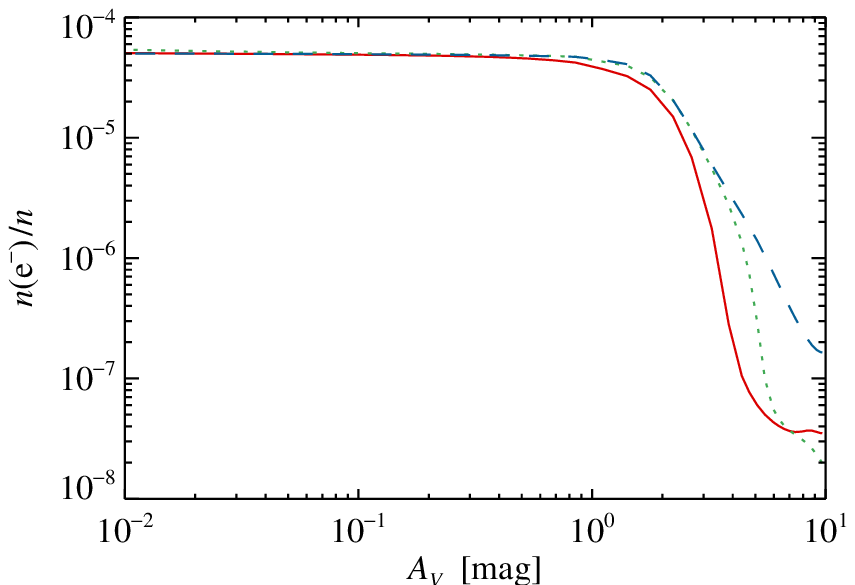}
 \caption{Electron fractional abundances for diffusion coefficients of $K=0$, $10^{21}$, and $10^{23}\kunit$ (red solid, green dotted, and blue dashed lines, respectively). The results for model \#1 are shown in the top plot and those for model \#2 in the bottom, both at $t=1$ Myr.}
 \label{Fig:Electron-Abundance}
\end{figure}

Fig.~\ref{Fig:Electron-Abundance} shows the fractional abundance of electrons as a function of visual extinction in both models, for values of $K$ from 0 (no diffusion) to $10^{23}\kunit$ (strong diffusion). The electron abundance in the cloud center rises by a factor of 4 when $K=10^{23}$, compared to models in which diffusion is absent. This is due to the influx of ions and electrons from the cloud surface, where species such as carbon are rapidly photoionized by the unattenuated flux of FUV photons.

\begin{figure}
 \plotone{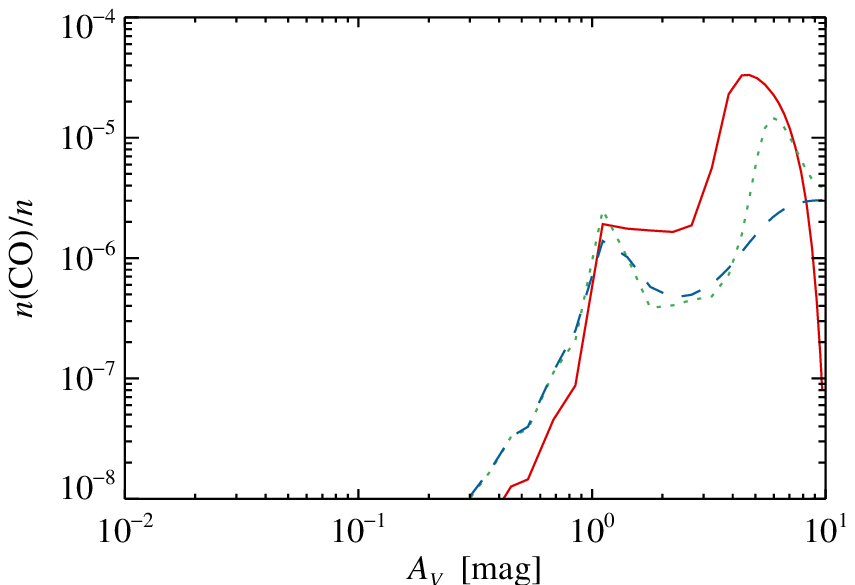}\\
 \plotone{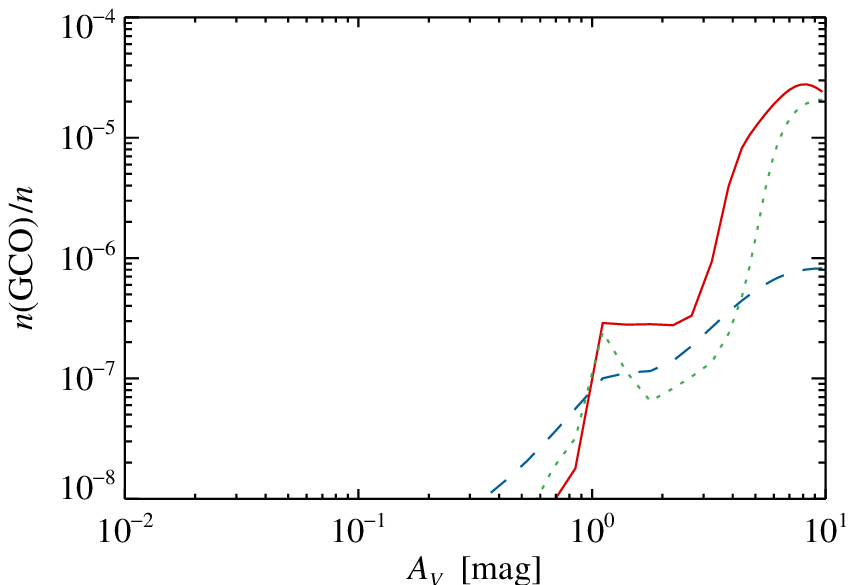}
 \caption{Fractional abundance of CO, both in the gas phase and on grain surfaces (referred to as GCO), for diffusion coefficients of $K=0$, $10^{21}$, and $10^{23}\kunit$ (red solid, green dotted, and blue dashed lines, respectively). The results shown are for model \#2 at $t=1$ Myr.}
 \label{Fig:Freeze-Out}
\end{figure}

Turbulent diffusion also acts to reduce the level of freeze-out in the cold dense cloud core. This is caused by movement of the grains to regions of lower extinction and density, where their mantles are released back into the gas, either by evaporation or photodesorption. As an example, gas-phase CO in the cloud center is over an order of magnitude more abundant in the $K=10^{23}\kunit$ models than in the models with no diffusion, whilst the grain mantle abundance of CO shows a corresponding drop (see Fig.~\ref{Fig:Freeze-Out}). Atomic oxygen, which rapidly destroys H$_3^+$, is 700 times more abundant in the $K=10^{23}$ model because its freeze-out onto grains is reduced.

\begin{figure}
 \plotone{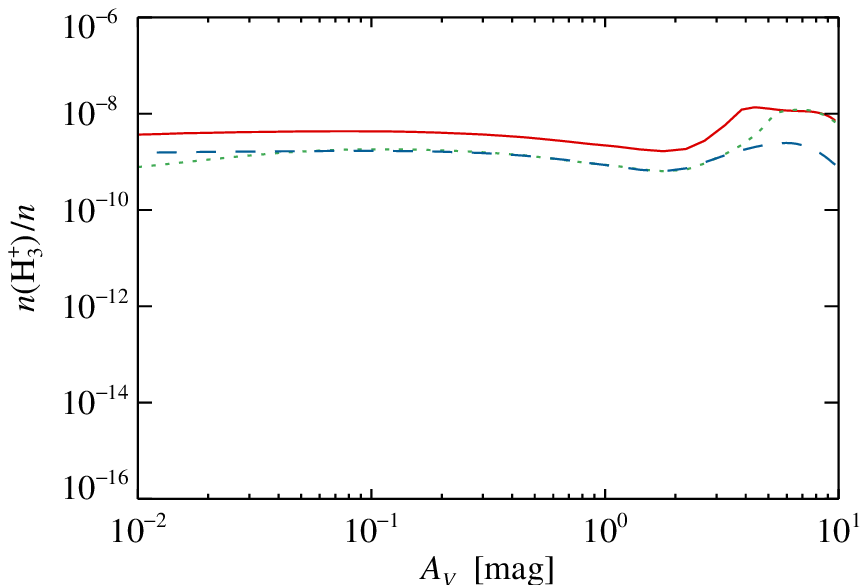}\\
 \plotone{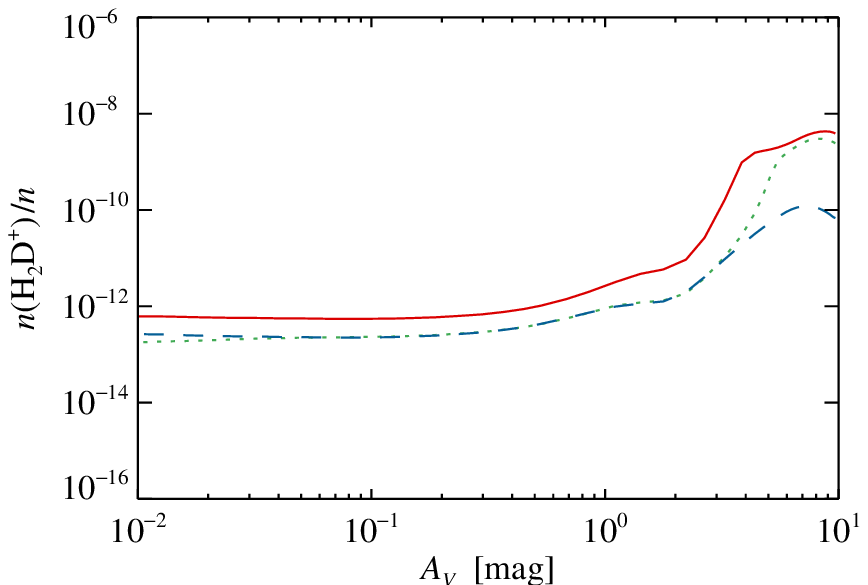}\\
 \plotone{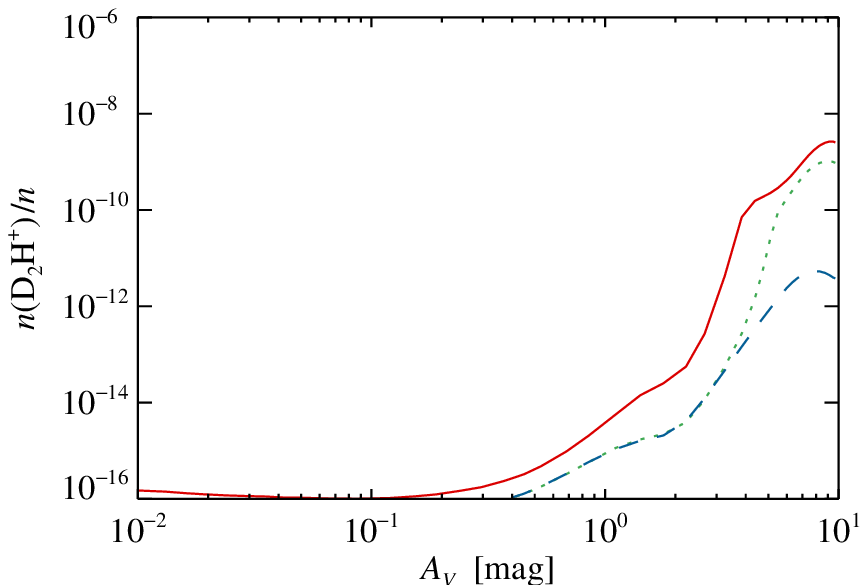}
 \caption{Fractional abundances of H$_3^+$, H$_2$D$^+$, and D$_2$H$^+$ for diffusion coefficients of $K=0$, $10^{21}$, and $10^{23}\kunit$ (red solid, green dotted, and blue dashed lines, respectively). The results shown are for model \#2 at $t=1$~Myr.}
 \label{Fig:Deuterated-Abundances}
\end{figure}

The combination of increased electron abundance and decreased freeze-out in the presence of turbulent diffusion leads to significantly lower abundances of H$_3^+$ and its deuterated analogs, compared to models with no diffusion. This in turn leads to lower abundances of the deuterated molecules that are formed from these ions. These trends are illustrated in Fig.~\ref{Fig:Deuterated-Abundances}, which shows the abundances of H$_3^+$, H$_2$D$^+$, and D$_2$H$^+$ dropping as the level of turbulent diffusion rises. Comparing models with $K=0$ and $K=10^{23}\kunit$, the abundances of H$_3^+$, H$_2$D$^+$, and D$_2$H$^+$ at the cloud center drop by factors of approximately 10, 30, and 200, respectively, when turbulent diffusion is active (for model \#1 parameters; the results for model \#2 show a similar trend).

Dissociative recombination of H$_2$D$^+$ and its multiply deuterated isotopologues is also the primary source of D atoms in the cloud center. As their abundances are lowered under the presence of turbulent diffusion, so too is the abundance of atomic deuterium---by a factor of $\sim$10$^4$ when $K=10^{23}~\kunit$. The lower abundance of D then inhibits the deuteration of molecules on grain surfaces, which proceeds via deuteron addition to mantle species. Since the high levels of deuteration observed toward low-luminosity protostars are often attributed to evaporation of mantle species following active grain surface chemistry \citep[e.g.,][]{Loinard2001, Roberts2003}, the role of turbulent diffusion in limiting grain surface deuteration may have implications for the degree of fractionation seen in the gas phase at later stages of protostellar evolution.

\begin{figure*}
 \plottwo{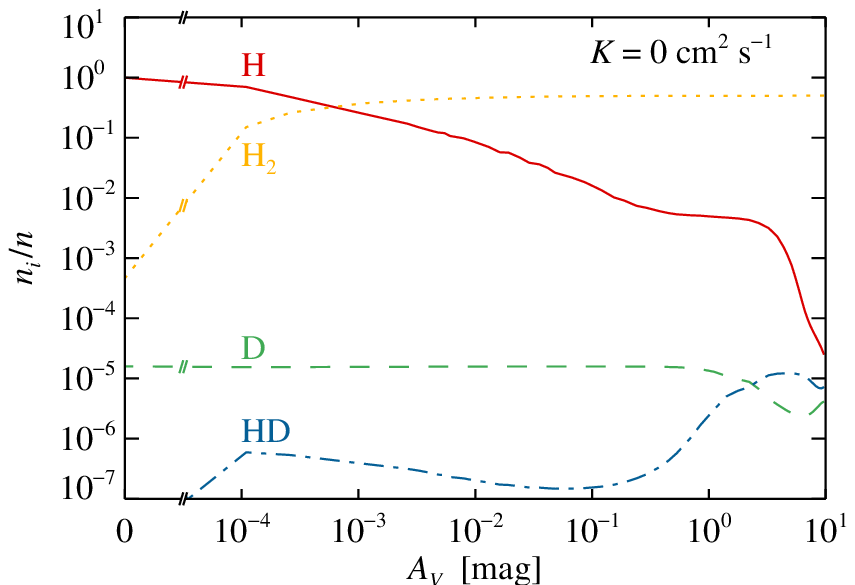}{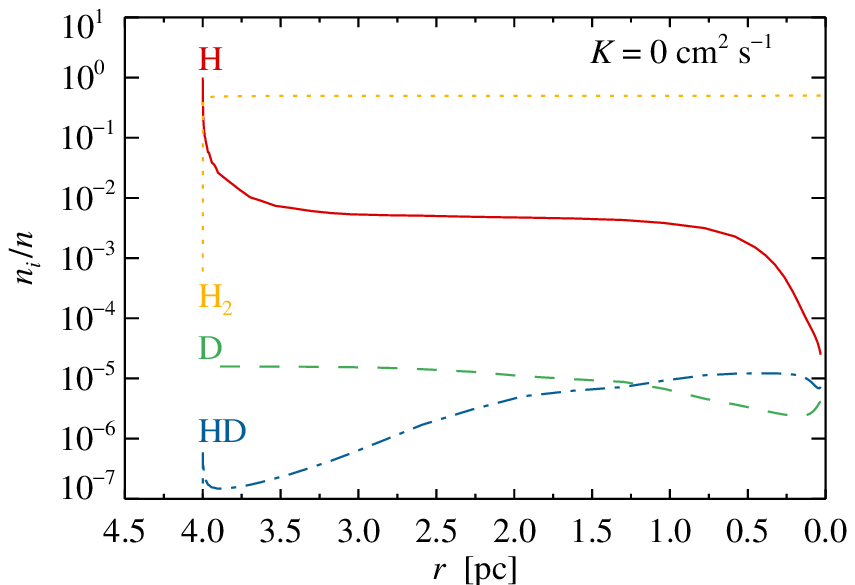}\\
 \plottwo{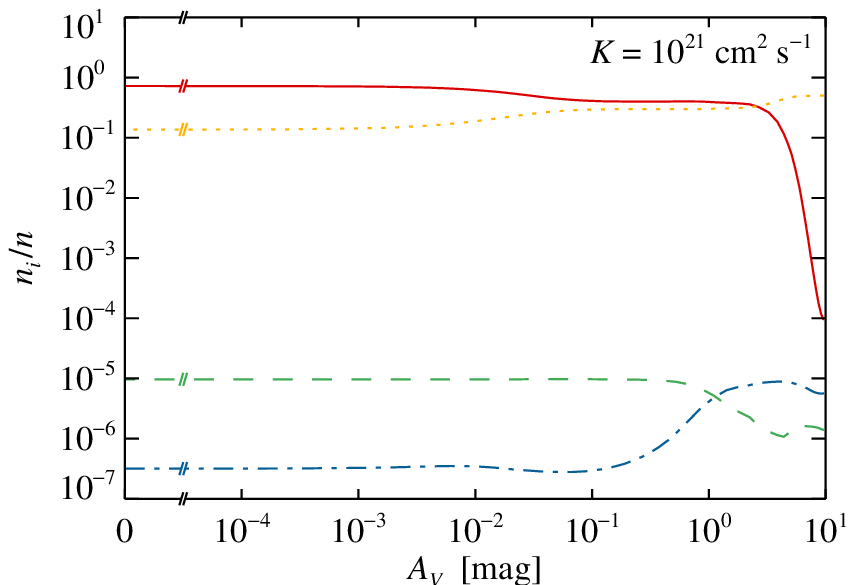}{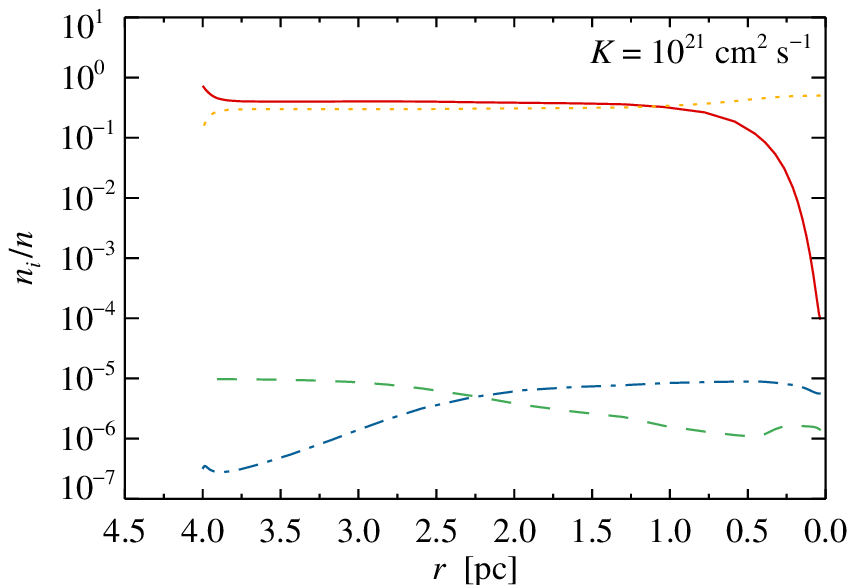}\\
 \plottwo{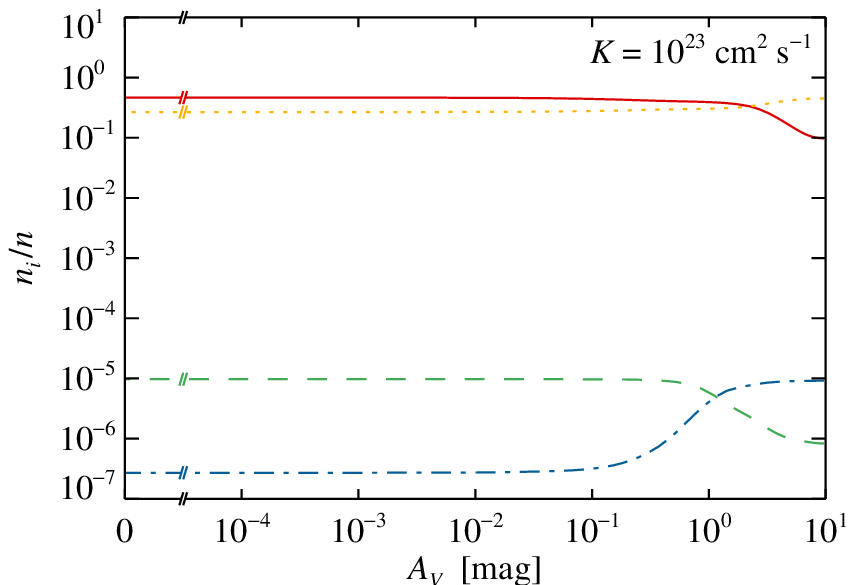}{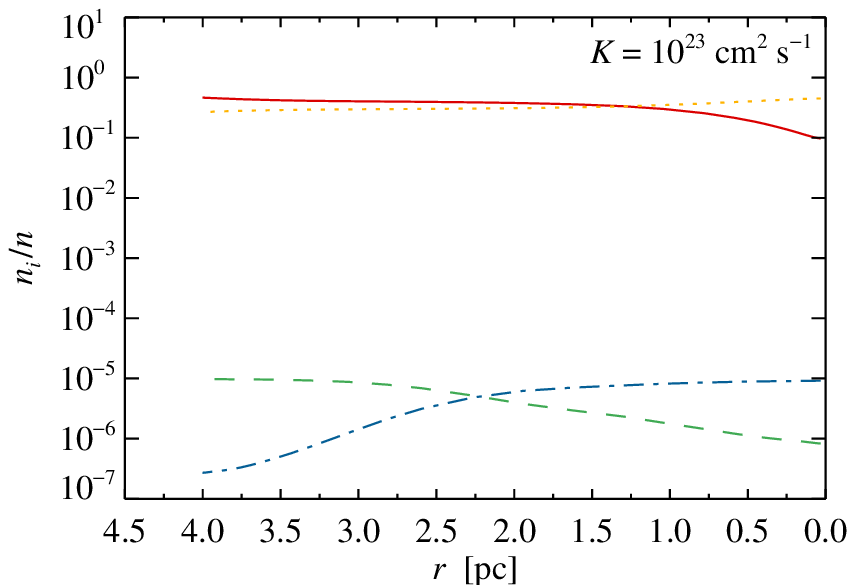}
 \caption{Fractional abundances of H, H$_2$, D, and HD (red solid, yellow dotted, green dashed, and blue dot-dashed lines, respectively) for diffusion coefficients of $K=0$, $10^{21}$, and $10^{23}\kunit$, from top to bottom. The results shown are for model \#2 at $t=1$ Myr.}
 \label{Fig:Transition-Regions}
\end{figure*}

\subsection{Cloud Envelope}\label{Cloud-Envelope}

At the cloud edge, diffusion affects the typical cloud chemistry by supplying fresh molecular gas to be photodissociated/photoionized by the incident FUV photons, the products of which are then carried into the cloud interior. This increases the depth at which the gas becomes molecular, shifting the locations of the H/H$_2$ and C$^+$/C/CO transitions that are characteristic features in models of cloud edges. However, the influence of turbulent diffusion is not uniform on all species. An interesting result of this is that the location of the H/H$_2$ transition can occur \textit{deeper} into the cloud than that of D/HD, despite the latter's weaker shielding against photodissociation. The reason for this lies in the importance of the abundance gradient in determining the rate of diffusion. Since the elemental abundance of hydrogen is orders of magnitude above that of deuterium, the abundance gradients of H and H$_2$ at the cloud edge, where the abundance of H$_2$ drops rapidly from 0.5 to $\sim$10$^{-3}$, are much steeper than those of D and HD. This is illustrated in Fig.~\ref{Fig:Transition-Regions}, where the abundances of H, H$_2$, D, and HD are shown as a function of visual extinction into the cloud for models with diffusion coefficients of $K=0$ to $10^{23}\kunit$. The inward flux of atomic hydrogen from the cloud surface, where it is produced by photodissociation of (primarily) H$_2$, is over four orders of magnitude greater than the inward flux of atomic deuterium. This has the effect of ``smearing out'' the otherwise sharp H/H$_2$ transition to such an extent that the depth at which H and H$_2$ have equal abundances increases from $A_V \sim 7$$\times$10$^{-4}$ ($K=0$) to $A_V \sim 2$ ($K=10^{23}$). In contrast, the equivalent position for D and HD moves only slightly, and in the opposite direction, from $A_V \sim 2$ to $A_V \sim 1$~mag.

Under typical interstellar conditions, molecular hydrogen is believed to form exclusively on grain surfaces. HD, however, is not only formed on grains, but can also form efficiently in the gas phase by the reaction
\begin{eqnarray}
 \rm{D^+ + H_2} & \rightarrow & \rm{H^+ + HD}. \nonumber
\end{eqnarray}
Indeed, in the warm diffuse cloud envelope, where the D$^+$ abundance is high due to charge exchange between D and H$^+$, and H$_2$ begins to become abundant, this gas-phase reaction becomes the dominant formation route for HD. The jump in HD abundance near the cloud edge ($A_V \sim 10^{-4}$~mag, $r \sim 4$~pc), seen in the top two plots of Fig.~\ref{Fig:Transition-Regions}, is due to the combination of high abundances of both D$^+$ (not shown) and H$_2$ at this point; nearer the cloud surface, H$_2$ is less abundant due to photodissociation, and further into the cloud, D$^+$ drops in abundance, due to the declining abundance of H and thus H$^+$, from which D$^+$ forms.

With the introduction of turbulent diffusion, the abundance gradient of D$^+$ is smoothed out, such that its abundance remains more constant across the cloud envelope and is a factor of $\sim$3 higher at $A_V=1$~mag when $K=10^{23}$. This, in turn, leads to increased gas-phase formation of HD at this cloud depth, resulting in a higher HD abundance and thus explaining the outward shift in the location of the D/HD transition. This change is relatively insensitive to the strength of turbulent diffusion; models with $K=10^{21}$ and $K=10^{23}\kunit$ produce nearly identical changes in the D$^+$ and HD abundances at this cloud depth.

\begin{figure}
 \plotone{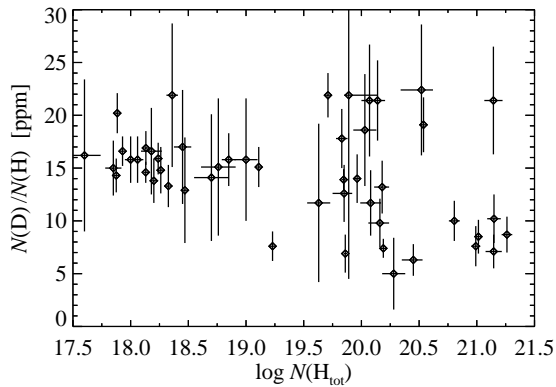}
 \caption{Observed galactic $\rm{[D]/[H]}$ ratios as a function of line-of-sight total hydrogen column density [$N(\mathrm{H_{tot}}) = N($H\,\textsc{i}$) + 2N(\rm{H_2})$]. See Table~\ref{Tab:Observed-D-to-H-Data} for more details and references.}
 \label{Fig:Observed-D-to-H-Ratio}
\end{figure}

Whilst the diffusive process acts to redistribute atomic species throughout the cloud, it also brings molecular material from the interior out to the envelope of the cloud. This allows certain deuterated species to attain higher abundances within the region of low extinction than can be produced by static cloud models, where the effects of photodissociation, lower density, and higher temperature inhibit the formation of such species. A notable example of this is N$_2$D$^+$, which shows an abundance enhancement of almost an order of magnitude when $K=10^{23}\kunit$, compared to the static cloud model (at $A_V\sim10^{-2}$~mag).

We note that observations of H$_3^+$ in diffuse clouds \citep[e.g.,][]{McCall2003, Indriolo2007} suggest that large low-energy cosmic ray ionization rates are needed to explain the observed high abundances. These rates are up to an order of magnitude greater than the standard rate considered in our model. Although it is beyond the scope of this paper to examine the possible influence of varying model parameters such as the cosmic ray ionization rate on the deuterium chemistry, it is worth mentioning that an order of magnitude increase in the rate would lead to higher H$_3^+$ and electron abundances in the model, which would likely reduce the level of deuteration that could be achieved. Since turbulent diffusion is effective at redistributing species between the cloud core and envelope, this would likely lead to an overall reduction in the level of deuteration throughout the cloud.


\begin{figure}
 \plotone{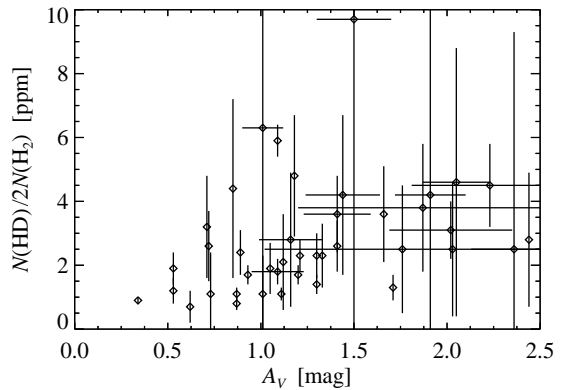}
 \caption{Observed galactic $\rm{[HD]/2[H_2]}$ ratios as a function of line-of-sight visual extinction [$A_V = R_V.E(B-V)$]. See Table~\ref{Tab:Observed-HD-to-H2-Data} for more details and references.}
 \label{Fig:Observed-HD-to-H2-Ratio}
\end{figure}

\begin{deluxetable}{lrcl}
 \tablewidth{70mm} 
 \tablecaption{Compilation of Observed Galactic $\rm{[D]/[H]}$ Ratios\label{Tab:Observed-D-to-H-Data}}
 \tablehead{
  \colhead{}       & \colhead{$d$}  & \colhead{} &
  \colhead{$\rm{[D]/[H]}$} \\
  \colhead{Target} & \colhead{(pc)} & \colhead{$\log N(\rm{H}\,\textsc{i})$} & \colhead{(ppm)}}
 \startdata
  Sirius         &    3 & 17.60 &    $16.2\pm7.2$ \\
  36 Oph         &    6 & 17.85 &    $15.0\pm2.5$ \\
  $\epsilon$ Eri &    3 & 17.88 &    $14.3\pm1.6$ \\
  31 Com         &   94 & 17.88 &    $20.2\pm1.9$ \\
  Hz 43          &   68 & 17.93 &    $16.6\pm1.4$ \\
  $\epsilon$ Ind &    4 & 18.00 &    $16.0\pm2.0$ \\
  Procyon        &    4 & 18.06 &    $16.0\pm2.0$ \\
  $\beta$ Cas    &   17 & 18.13 &    $16.9\pm1.6$ \\
  HR 1099        &   29 & 18.13 &    $14.6\pm1.0$ \\
  G191--B2B      &   69 & 18.18 &    $16.6\pm4.1$ \\
  $\beta$ CMa    &  153 & 18.20 & \multicolumn{1}{c}{$\ge$16} \\
  $\sigma$ Gem   &   37 & 18.20 &    $13.8\pm2.1$ \\
  Capella        &   13 & 18.24 &    $15.9\pm1.5$ \\
  $\beta$ Gem    &   10 & 18.26 &    $14.8\pm2.2$ \\
  $\alpha$ Tri   &   20 & 18.33 &    $13.3\pm2.0$ \\
  $\beta$ Cet    &   29 & 18.36 &    $21.9\pm6.8$ \\
  $\lambda$ And  &   26 & 18.45 &    $17.0\pm5.0$ \\
  Feige 24       &   74 & 18.47 &    $13.0\pm5.0$ \\
  WD 0621--376   &   78 & 18.70 &    $14.1\pm6.0$ \\
  WD 2211--495   &   53 & 18.76 &    $15.1\pm6.5$ \\
  WD 1634--573   &   37 & 18.85 &    $15.8\pm2.5$ \\
  $\alpha$ Vir   &   80 & 19.00 &    $15.8\pm8.0$ \\
  GD 246         &   79 & 19.11 &    $15.1\pm1.9$ \\
  $\lambda$ Sco  &  216 & 19.23 & \phn$7.6\pm1.8$ \\
  $\beta$ Cen    &  161 & 19.63 &    $11.7\pm7.5$ \\
  $\gamma$ Vel   &  258 & 19.71 &    $21.8\pm2.1$ \\
  $\alpha$ Cru   &   98 & 19.85 &    $12.6\pm3.6$ \\
  BD $+$28 4211  &  104 & 19.85 &    $13.9\pm1.0$ \\
  Lan 23         &  122 & 19.89 &    $21.9\pm10.8$\\
  $\mu$ Col      &  400 & 19.86 & \phn$6.9\pm6.9$ \\
  $\zeta$ Pup    &  429 & 19.96 &    $14.0\pm2.3$ \\
  TD1 32709      &  520 & 20.03 &    $18.6\pm5.3$ \\
  WD 1034$+$001  &  155 & 20.07 &    $21.4\pm5.3$ \\
  BD $+$39 3226  &  290 & 20.08 &    $11.7\pm3.1$ \\
  Feige 110      &  179 & 20.14 &    $21.4\pm5.7$ \\
  $\gamma$ Cas   &  188 & 20.16 & \phn$9.8\pm2.7$ \\
  $\iota$ Ori    &  407 & 20.15 &    $14.1\pm2.8$ \\
  $\delta$ Ori   &  281 & 20.19 & \phn$7.4\pm1.2$ \\
  $\theta$ Car   &  135 & 20.28 & \phn$5.0\pm3.4$ \\
  $\epsilon$ Ori &  412 & 20.45 & \phn$6.3\pm1.8$ \\
  PG 0038$+$199  &  297 & 20.48 &    $19.1\pm2.6$ \\
  LSE 44         &  554 & 20.52 &    $22.4\pm9.0$ \\
  JL 9           &  590 & 20.78 &    $10.0\pm1.9$ \\
  HD 195965      &  794 & 20.95 & \phn$8.5\pm1.6$ \\
  LSS 1274       &  580 & 20.98 & \phn$7.6\pm1.9$ \\
  HD 191877      & 2200 & 21.05 & \phn$7.8\pm2.4$ \\
  HD 41161       & 1253 & 21.08 &    $21.6\pm5.1$ \\
  HD 53975       & 1318 & 21.14 &    $10.2\pm2.3$ \\
  HD 90087       & 2740 & 21.22 & \phn$8.7\pm1.7$ \\[-2mm]
 \enddata
 \tablecomments{Data taken from \citet{Linsky2006}, \citet{Oliveira2006}, and \citet{Ellison2007}.}
\end{deluxetable}

\begin{deluxetable}{lrcccl}
 \tablecaption{Compilation of Observed Galactic $\rm{[HD]/2[H_2]}$ Ratios\label{Tab:Observed-HD-to-H2-Data}}
 \tablehead{
  \colhead{}       & \colhead{$d$}  & \colhead{$A_V$} & \colhead{}                       & \colhead{}       & \colhead{$\rm{[HD]/2[H_2]}$} \\
  \colhead{Target} & \colhead{(pc)} & \colhead{(mag)} & \colhead{$\log N(\rm{H_{tot}})$} & \colhead{$f(\rm{H_2})$} & \colhead{(ppm)}}
 \startdata
  HD 12323  & 3900 & 0.72 & 21.29 & 0.21 & $2.6\pm1.1$ \\
  HD 15558  & 2187 & 2.44 & 21.69 & 0.32 & $2.8\pm2.1$ \\
  HD 24534  &  398 & 2.05 & 21.34 & 0.76 & $4.6\pm4.2$ \\
  HD 27778  &  220 & 1.01 & 21.34 & 0.56 & $6.3\pm16.0$\\
  HD 45314  &  799 & 2.03 & 21.28 & 0.42 & $2.5\pm2.1$ \\
  HD 53367  &  247 & 1.76 & 21.65 & 0.51 & $2.5\pm2.0$ \\
  HD 73882  &  925 & 2.36 & 21.58 & 0.67 & $2.5\pm6.8$ \\
  HD 74920  & 1497 & 1.09 & 21.25 & 0.21 & $5.9\pm0.5$ \\
  HD 90087  & 2716 & 0.93 & 21.18 & 0.08 & $1.7\pm0.3$ \\
  HD 91651  & 3500 & 1.01 & 21.16 & 0.02 & $1.1\pm1.2$ \\
  HD 91824  & 4000 & 0.87 & 21.19 & 0.09 & $1.1\pm0.2$ \\
  HD 93204  & 2630 & 1.30 & 21.43 & 0.04 & $1.4\pm0.3$ \\
  HD 93205  & 2600 & 1.20 & 21.35 & 0.05 & $1.7\pm0.3$ \\
  HD 93206  & 2512 & 1.21 & 21.35 & 0.03 & $2.3\pm0.5$ \\
  HD 93222  & 2900 & 1.71 & 21.55 & 0.03 & $1.3\pm0.4$ \\
  HD 94493  & 3327 & 0.62 & 21.17 & 0.18 & $0.7\pm0.5$ \\
  HD 101131 &  709 & 1.05 &\nodata&\nodata&$1.9\pm0.8$ \\
  HD 101190 & 2399 & 0.89 & 21.29 & 0.27 & $2.4\pm0.7$ \\
  HD 101413 & 2399 & 1.12 & 21.34 & 0.22 & $2.1\pm1.5$ \\
  HD 101436 & 2399 & 1.18 & 21.34 & 0.22 & $4.8\pm1.9$ \\
  HD 104705 & 3898 & 0.73 & 21.16 & 0.12 & $1.1\pm1.3$ \\
  HD 110432 &  301 & 2.02 & 21.20 & 0.55 & $3.1\pm0.9$ \\
  HD 116852 & 4760 & 0.53 & 21.01 & 0.12 & $1.2\pm0.4$ \\
  HD 147888 &  136 & 1.91 & 21.71 & 0.18 & $4.2\pm14.1$\\
  HD 148422 & 8836 & 0.85 & 21.23 & 0.16 & $4.4\pm2.8$ \\
  HD 149404 & 1380 & 2.23 & 21.57 & 0.33 & $4.5\pm1.3$ \\
  HD 152233 & 1905 & 1.33 & 21.37 & 0.17 & $2.3\pm1.0$ \\
  HD 152248 & 1758 & 1.66 &\nodata&\nodata&$3.6\pm1.5$ \\
  HD 152723 & 1905 & 1.41 & 21.49 & 0.13 & $2.6\pm0.8$ \\
  HD 161807 &  383 & 0.34 &\nodata&\nodata&$0.9\pm0.1$ \\
  HD 177989 & 4909 & 0.71 & 21.06 & 0.23 & $3.2\pm1.6$ \\
  HD 185418 &  950 & 1.16 & 21.56 & 0.47 & $2.8\pm2.1$ \\
  HD 192639 & 1100 & 1.87 & 21.47 & 0.32 & $3.8\pm2.0$ \\
  HD 199579 &  794 & 1.09 & 21.25 & 0.38 & $1.8\pm0.4$ \\
  HD 201345 & 1907 & 0.53 & 20.91 & 0.03 & $1.9\pm0.5$ \\
  HD 206267 &  850 & 1.41 & 21.54 & 0.42 & $3.6\pm1.2$ \\
  HD 207198 &  832 & 1.50 & 21.55 & 0.38 & $9.7\pm10.6$\\
  HD 207538 &  832 & 1.44 & 21.58 & 0.43 & $4.2\pm2.5$ \\
  HD 224151 & 1355 & 1.11 & 21.45 & 0.26 & $1.1\pm0.2$ \\
  HD 303308 & 2630 & 1.30 & 21.50 & 0.11 & $2.3\pm0.7$ \\
  HD 308813 & 2398 & 0.87 & 21.26 & 0.22 & $0.8\pm0.2$ \\[-2mm]
 \enddata
 \tablecomments{Data taken from \citet{Snow2008}, with additional extinction and $N(\rm{H_2})$ data taken from \citet{Rachford2002, Rachford2009}. For targets without $R_V$ data, a typical value of 3.1 is adopted when calculating the visual extinction [$A_V = R_V.E(B-V)$].}
\end{deluxetable}

\section{Implications for the Observed D/H Ratio}\label{D-to-H-Ratio}

The galactic atomic D/H ratio has been studied extensively by observing absorption lines of \ion{D}{1} and \ion{H}{1} toward background stars in the local ISM. In the past, the lines of sight were restricted to nearby stars ($d \lesssim 200$~pc) by the sensitivity limits of the {\it Copernicus} and {\it Hubble Space Telescope} instruments, but with the launch of the {\it Far Ultraviolet Spectroscopic Explorer} ({\it FUSE}) satellite, absorption studies along sightlines to stars several kpc away became possible. Fig.~\ref{Fig:Observed-D-to-H-Ratio} and Table~\ref{Tab:Observed-D-to-H-Data} show a compilation of the observed galactic D/H ratios, taken primarily from the compilation of \citet{Linsky2006}, but with more recent measurements added. The scatter in observed values is significant, with some measurements up to factor of 2 above or below the mean value.

More recently, the {\it molecular} D/H ratio, defined as [HD]/2[H$_2$], has begun to be examined in more detail, since {\it FUSE} allows sightlines containing significant fractions of molecular material to be probed. Fig.~\ref{Fig:Observed-HD-to-H2-Ratio} and Table~\ref{Tab:Observed-HD-to-H2-Data} show a compilation of the observed galactic [HD]/2[H$_2$] ratios, taken primarily from \citet{Snow2008}, but updated with more recent data.

Let us now assume that our model cloud intersects a line of sight toward a background star and that we measure the column densities of \ion{H}{1} and \ion{D}{1} seen in absorption through the cloud. Given the abundance profiles of H, H$_2$, D, and HD discussed in the previous section, what atomic D/H ratios might we expect at different points within the cloud? If the cloud consisted purely of atomic gas, measurements along any line of sight through it would yield the true elemental D/H ratio (i.e., that specified as an input to the model). However, if we consider a model cloud containing cold, dense material, much of the hydrogen and deuterium will be contained in molecules, either in the gas phase or on grain mantles. In this case, the measured atomic D/H ratio will no longer necessarily reflect the true elemental D/H ratio. Instead, it will reflect the \textit{remaining} amount of hydrogen and deuterium in atomic form along a particular sightline, and will no longer ``see'' the deuterium and hydrogen contained in molecules. At this point, the measured D/H ratio becomes dependent upon the various chemical reactions that form and remove atomic deuterium and hydrogen. Due to the mass difference between these isotopologues, the formation and destruction rates for H- and D-bearing species will differ at low temperatures as a result of fractionation, and the fraction of each that remains in atomic form will strongly depend on the density and temperature of the cloud. Furthermore, if the cloud is immersed in an interstellar radiation field, the photodissociation of molecules by the impinging UV photons will also affect the amount of atomic deuterium and hydrogen present in the cloud envelope. The self-shielding of H$_2$ means that, as extinction into the cloud increases, its photodissociation rate falls off more rapidly than that of HD \citep[for a detailed discussion, see][]{LePetit2002}. This difference in shielding results in a significant fraction of atomic deuterium being maintained to greater cloud depth, whereas atomic hydrogen rapidly transitions to molecular form. Model clouds containing density and temperature gradients, or being exposed to an external radiation field, will therefore yield depth-dependent D/H ratios.

\begin{figure}
 \plotone{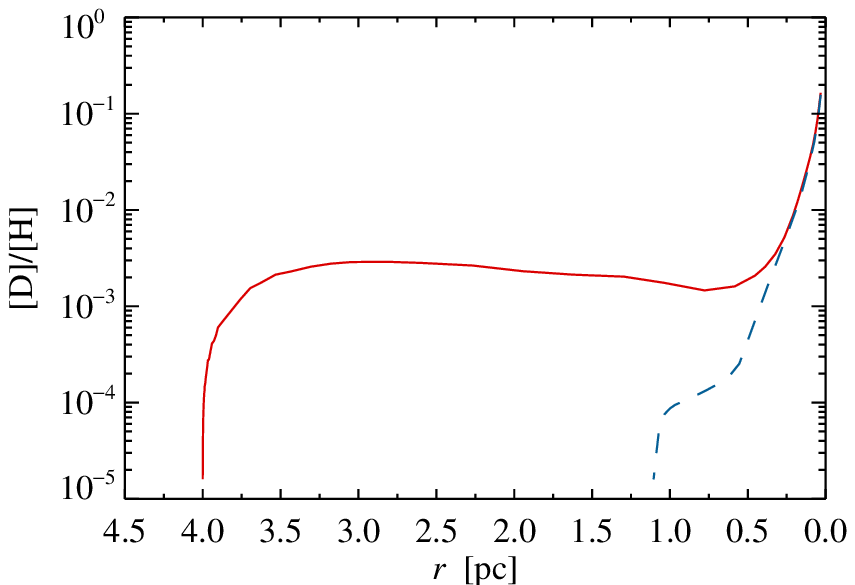}\\
 \plotone{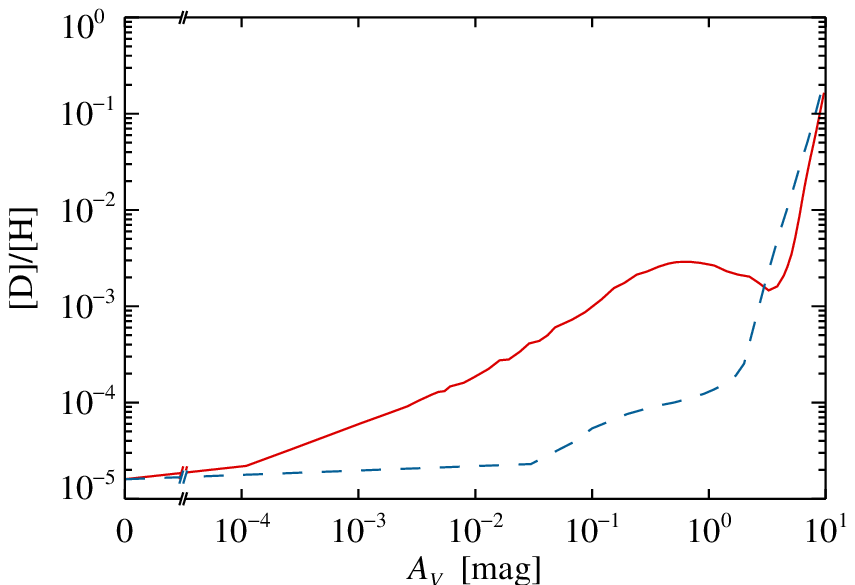}
 \caption{The atomic D/H ratio for static (i.e., no diffusion) versions of model \#1 (blue, dashed) and \#2 (red, solid), both at $t=1$ Myr. The top plot shows the radial dependence of the D/H ratio and the bottom plot shows its variation as a function of visual extinction into the cloud.}
 \label{Fig:Static-D-to-H-Ratio}
\end{figure}

Fig.~\ref{Fig:Static-D-to-H-Ratio} shows the atomic D/H ratio determined from the ratio of abundances, $\rm{[D]/[H]}$, as a function of visual extinction and radius in our static cloud model (i.e., with no diffusion). It is clear that the chemical processes discussed above can lead to a dramatic variation of the D/H ratio within such a cloud, with values spanning 4 orders of magnitude across its radius. With increasing depth (or visual extinction) into the cloud, the D/H ratio is seen to rise sharply, as hydrogen rapidly transitions to H$_2$ due to its effective shielding, whilst the density of atomic deuterium remains high, with a more gradual transition from D to HD, owing to its poor shielding. Going deeper into the cloud, the atomic D/H ratio levels off as hydrogen becomes predominantly molecular and deuterium remains in atomic form, before the ratio begins to drop again, as deuterium, too, moves to molecular form (primarily HD). Finally, in the dark cloud interior ($A_V > 1$~mag), the D/H ratio starts to rise again as deuteration of H$_3^+$ and its subsequent dissociative recombination preferentially leads to the formation of D, rather than H, atoms in the cold dense core. Fractionation is so effective in the core that the atomic D/H ratio exceeds 0.1, a factor of 10$^4$ increase over the cosmic D/H ratio.

As we have seen in \S\ref{Results}, introducing turbulent mixing to the model changes the amount of atomic deuterium and hydrogen present in the cloud by causing these atomic species to diffuse along their abundance gradients. Changes in the abundance of H and D due to turbulent diffusion lead to a corresponding change in the atomic D/H ratio. To examine the effect of increasing turbulent diffusion on the D/H ratio in the outer regions of the cloud, we focus our attention on the results of model \#2, which includes a warm diffuse envelope that better represents the conditions at the edges of molecular clouds.

In \S\ref{Cloud-Envelope} we showed that the abundances of atomic deuterium and hydrogen tend to drop at the cloud edge as the level of turbulence increases. This is due to their inward diffusion along the abundance gradients that are created by the H/H$_2$ and D/HD transitions. However, the strength of this diffusion is not the same for both species, and the abundance profile of atomic hydrogen is changed more significantly than that of atomic deuterium. This leads to changes in the atomic D/H ratio as the level of turbulent diffusion is increased.

At $K=10^{18}\kunit$, the D/H ratio at the cloud edge drops from the elemental value (16 ppm) to about 10 ppm. This is due to the drop in atomic deuterium abundance whilst the abundance of hydrogen remains relatively unchanged. As the level of turbulence increases, the inward diffusion of atomic hydrogen becomes more active and its abundance at the cloud edge begins to drop. Atomic deuterium is less effected, since its abundance gradient is smaller and occurs deeper into the cloud. The net result is that the D/H ratio begins to rise again, reaching 21 ppm when $K=10^{23}\kunit$, the maximum diffusion coefficient we consider here.

Deeper into the cloud, the introduction of turbulent diffusion smoothes out the abundance profiles of atomic deuterium and hydrogen, leading to less variation in the atomic D/H profile with cloud depth. In the cloud envelope ($A_V < 5$~mag), even a low level of turbulent mixing can dramatically reduce the variation of the D/H ratio, from over two orders of magnitude when no turbulent mixing is present to just a factor of $\sim$4 when $K=10^{18}\kunit$. In the cloud core ($A_V \sim 10$~mag), turbulent diffusion has less effect on the atomic D/H ratio, which only drops by a factor of 10 when $K=10^{21}\kunit$, owing to the greater influx of atomic hydrogen and the reduced abundance of deuterated H$_3^+$, the main source of D atoms in the cold dense core. Increasing the diffusion coefficient to $K=10^{23}\kunit$, however, causes the D/H ratio in the core to drop to $\approx$9 ppm, a factor of 10$^4$ reduction compared to the static model with no diffusion. Strong turbulence can therefore have a bigger impact on the atomic D/H ratio in the core. At this level of turbulence, diffusive mixing is so effective that the abundance of atomic hydrogen is fairly uniform across the cloud, varying by less than a factor of 5, whereas, in the static model, it is nearly 10$^5$ lower at the core than at the edge.

The effect of varying the strength of turbulent diffusion in the model is therefore to produce a range of atomic D/H ratios at the cloud edge, with values spanning $\sim$10--20 ppm. In the cloud interior, increasing turbulence causes a more significant drop in the atomic D/H ratio, by up to 4 orders of magnitude in the case of strong diffusion ($K=10^{23}\kunit$).

\begin{figure}
 \plotone{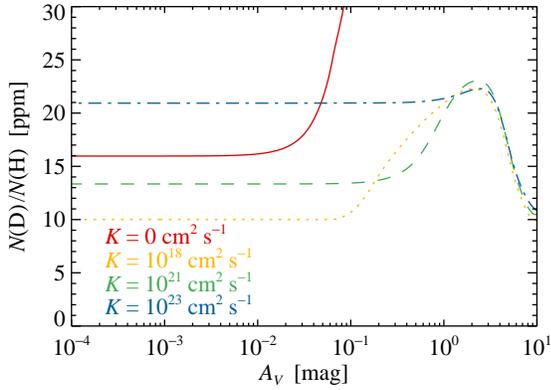}
 \caption{The integrated line-of-sight atomic D/H ratio through the model cloud, i.e., $N(\rm{D})/N(\rm{H})$, along sightlines of increasing extinction through the cloud, for diffusion coefficients of $K=0$, $10^{18}$, $10^{21}$, and $10^{23}\kunit$ (red solid, yellow dotted, green dashed, and blue dot-dashed lines, respectively). The results shown are for model \#2 at $t=1$ Myr.}
 \label{Fig:D-to-H-Column}
\end{figure}

If the lines of sight along which the galactic atomic D/H ratio have been measured are assumed to intersect diffuse gas at the edges of molecular clouds, the variation of the observed column densities of \ion{D}{1} and \ion{H}{1} (and therefore the inferred D/H ratio) will be dependent upon the degree of turbulent mixing within the clouds. Fig.~\ref{Fig:D-to-H-Column} shows the atomic D/H ratio that would be observed along a line of sight through a spherically symmetric model cloud, as a function of visual extinction through the cloud, for the case of no turbulent diffusion ($K=0\kunit$) up to strong diffusion ($K=10^{23}\kunit$). It can be seen that, even for sightlines that intersect the cloud near its edge, the atomic D/H ratio can change by over a factor of two. If clouds exhibiting varying degrees of turbulent mixing are assumed to exist along different sightlines (as is almost certainly the case), this could help to account for the observed range of atomic D/H ratios.

Other mechanisms invoked to explain the observed variation in the atomic D/H ratio (in particular, depletion of deuterium onto dust grains or galactic infall of nearly pristine material) are expected to also produce variations in the observed metal abundances. In the case of the deuterium depletion model, the observed abundances of refractory metals are predicted to vary in line with that of deuterium, since they exhibit similar depletion onto grains. Conversely, increases in the D/H ratio produced by unmixed infalling low metallicity material should be accompanied by decreased metal abundances. However, our turbulent diffusion models show only minor variations ($<$5\%) in the metal abundances (\ion{O}{1}, \ion{Fe}{2}, and \ion{Mg}{2}) that would be observed at the cloud surface over the full range of turbulent diffusion coefficients considered. This suggests that there would be only a weak correlation between the D/H ratio and the metal abundances if turbulent diffusion were the sole cause of variation in the atomic D/H ratio.

\begin{figure}
 \plotone{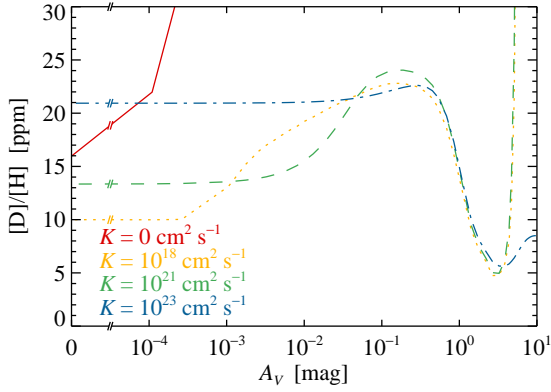}
 \caption{The atomic D/H ratio as a function of visual extinction into the cloud for diffusion coefficients of $K=0$, $10^{18}$, $10^{21}$, and $10^{23}\kunit$ (red solid, yellow dotted, green dashed, and blue dot-dashed lines, respectively). The results shown are for model \#2 at $t=1$ Myr.}
 \label{Fig:D-to-H-Ratio}
\end{figure}

\begin{figure}
 \plotone{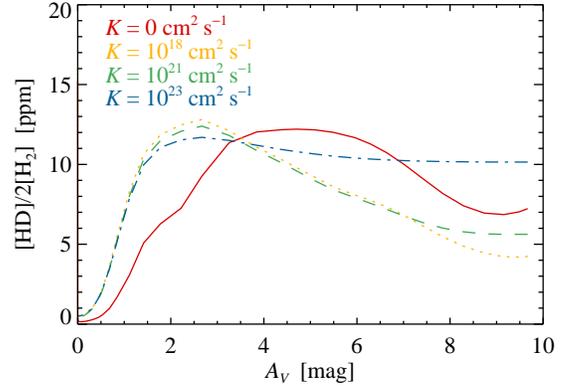}
 \caption{The molecular [HD]/2[H$_2$] ratio as a function of visual extinction into the cloud for diffusion coefficients of $K=0$, $10^{18}$, $10^{21}$, and $10^{23}\kunit$ (red solid, yellow dotted, green dashed, and blue dot-dashed lines, respectively). The results shown are for model \#2 at $t=1$ Myr.}
 \label{Fig:HD-to-H2-Ratio}
\end{figure}

Fig.~\ref{Fig:HD-to-H2-Ratio} shows the molecular D/H ratio determined from the ratio of abundances, [HD]/2[H$_2$], as a function of visual extinction into the cloud. In comparison to the atomic D/H ratio, the molecular equivalent shows significantly less variation across the cloud (2 orders of magnitude, instead of 4 orders of magnitude). In the static cloud model, the molecular D/H ratio attains the elemental D/H ratio (16~ppm) at the cloud edge, since the same unattenuated photodissociation rate is assumed for H$_2$ and HD, before dropping rapidly to $<$1~ppm when $A_V \gtrsim 10^{-4}$~mag, as self-shielding by H$_2$ causes its abundance to rise sharply. Deeper into the cloud ($A_V \sim 0.1$~mag), the abundance of HD begins to rise, as the the density increases and its rate of photodissociation starts to drop off, whilst the abundance of H$_2$ remains fairly constant. This leads to a steady increase in the [HD]/2[H$_2$] ratio, which peaks at about 12~ppm. In the cloud core, efficient fractionation transfers the deuterium from HD to other molecules (and to atomic D), and the molecular D/H ratio begins to fall again, reaching 7~ppm at the center.

\begin{figure}
 \plotone{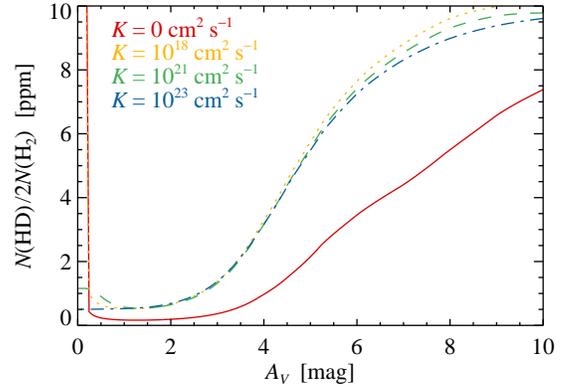}
 \caption{The integrated line-of-sight molecular D/H ratio through the model cloud, i.e., $N(\rm{HD})/2N(\rm{H_2})$, along sightlines of increasing extinction through the cloud, for diffusion coefficients of $K=0$, $10^{18}$, $10^{21}$, and $10^{23}\kunit$ (red solid, yellow dotted, green dashed, and blue dot-dashed lines, respectively). The results shown are for model \#2 at $t=1$ Myr.}
 \label{Fig:HD-to-H2-Column}
\end{figure}

The influence of turbulent diffusion on the molecular D/H ratio serves to lower the ratio at the cloud edge (to $\sim$1~ppm when $K=10^{21}\kunit$), since the mixing is more effective at bringing H$_2$ to the cloud surface than HD. The outward shift of the D/HD transition caused by turbulent diffusion (see \S\ref{Cloud-Envelope}) means that [HD]/2[H$_2$] begins to rise at lower $A_V$, and the ratio peaks at 2--3~mag, compared to 4--5~mag in the static cloud model. This shift is insensitive to the strength of turbulent diffusion. Changes in the level of fractionation brought about by turbulent mixing affect the amount of deuterium transferred from HD to other molecules and lead to molecular D/H ratios varying between 5 and 10~ppm at the cloud center.

Molecular D/H ratios determined from line-of-sight column densities through the model cloud, $N(\rm{HD})/2N(\rm{H_2})$, are shown as a function of line-of-sight visual extinction in Fig.~\ref{Fig:HD-to-H2-Column} for turbulent diffusion coefficients of 0 to 10$^{23}\kunit$. Comparing this figure to the observed molecular D/H ratios in Fig.~\ref{Fig:Observed-HD-to-H2-Ratio}, which have been measured for sightlines showing up to 2.5~mag of extinction, it is clear that our models underpredict the ratio by a factor of 3--4. However, models including some degree of turbulent mixing give the best agreement, with generally higher molecular D/H ratios over the observed range of $A_V$. It has been noted previously \citep{LePetit2002, Lacour2005} that models fail to match the observed values unless the bulk of the absorbing gas is assumed to be in a single, dense component. Since our model cloud possesses a fairly large envelope of diffuse and warm material, which dominates the line-of-sight column densities at low $A_V$, this could account for the difference. If the [HD]/2[H$_2$] abundance ratios in Fig.~\ref{Fig:HD-to-H2-Ratio} are instead compared to the observations, somewhat better agreement is found at low visual extinction ($<$1~mag). The agreement worsens at higher extinction, though this may be due in part to the larger uncertainties on the observed values at higher $A_V$.


\section{Summary}\label{Summary}

We find that turbulent diffusion affects the cloud chemistry in the following ways:
\begin{itemize}
\item The ionization fraction is enhanced in the cloud center, due to an influx of ions and electrons produced by photoionization at the surface.
\item Freeze-out of abundant gas-phase species, such as CO, is reduced, as grain mantles are photodesorbed and/or evaporated when they are brought to the cloud surface.
\item Abundances of H$_2$D$^+$, D$_2$H$^+$, and D$_3^+$ are reduced in the cloud center, due to the higher ionization fraction and CO abundance.
\item The H/H$_2$ transition occurs significantly deeper into the cloud, due to rapid inward transport of atomic hydrogen along the sharp abundance gradient caused by self-shielding of molecular hydrogen.
\item The D/HD transition is affected less by turbulent mixing, as the abundance gradients of D and HD are more gradual than those of H and H$_2$ (due to the lower elemental deuterium abundance and less effective shielding of HD).
\end{itemize}
These effects lead to generally lower abundances of deuterated molecules and a corresponding drop in fractionation ratios. We find that turbulent diffusion causes atomic hydrogen to remain abundant deep into molecular clouds, whilst inhibiting the pathways that normally allow significant deuteration in the cold dense cores of these clouds. This is reflected in the atomic D/H ratio, which drops dramatically from $\sim$10$^{-1}$ in the cloud center when $K=0\kunit$ (i.e., no diffusion), to $\sim$10$^{-5}$ when $K=10^{23}\kunit$. Close to the cloud edge, the atomic D/H ratio varies by over a factor of 2 as the level of turbulent mixing increases, with values spanning $\sim$10--20 ppm, in agreement with the range of observed values. The impact of turbulent mixing on the molecular D/H ratio, [HD]/2[H$_2$], is less significant, but nevertheless shows some dependence on the degree of turbulent mixing.

Based on these results, we propose a new process to help explain the significant scatter in the observed galactic atomic D/H ratio. If part of the intervening gas resides in the envelope of a molecular cloud, we expect the column densities of \ion{H}{1} and \ion{D}{1} near the cloud edge to be dependent on the degree of turbulent mixing within the cloud, leading to variations in the D/H ratio that would be observed.

In reality, various mechanisms are likely to be working in combination to produce the observed range of D/H ratios. Many of the parameters that govern the relative contribution of each process, such as the initial abundances, dust composition, turbulent diffusion coefficient, and ionization fraction, are hard to determine with any precision, making it difficult to disentangle the various processes involved, or to say difinitively if one process is dominating. The turbulent diffusion mechanism proposed here could be tested by searching for a correlation between observed D/H ratios and a proxy for the degree of turbulence along each line of sight, such as the turbulent linewidths of the absorption lines, though the analysis would be complicated by additional uncertainties due to overlapping absorption components and bulk cloud motions.

With the exception of galactic infall of low metallicity material, all mechanisms that have so far been proposed to explain the variation in atomic D/H ratio result in lower values. Therefore, the appropriate value to adopt would be the largest observed, i.e., $\sim$23 ppm in the Milky Way \citep[e.g.,][]{Linsky2006}. The turbulent mixing of atomic and molecular gas at the edges of molecular clouds discussed here is an alternative mechanism that can explain the scatter in atomic D/H ratios observed beyond the Local Bubble. However, in this case the observed D/H ratio can be either decreased or increased compared to the ``true'' elemental ratio, depending on the magnitude of the turbulent diffusion coefficient.


\acknowledgments

We thank W.~D.~Langer for helpful discussions and the referee for constructive comments which helped to improve an earlier draft of this paper. This research has been supported by the National Science Foundation grant AST-0838261 to the Caltech Submillimeter Observatory. Part of this research was carried out at the Jet Propulsion Laboratory, California Institute of Technology, under a contract with the National Aeronautics and Space Administration.



\begin{thebibliography}{}

\bibitem[{{Aikawa} {et~al.}(2005){Aikawa}, {Herbst}, {Roberts}, \&
  {Caselli}}]{Aikawa2005}
{Aikawa}, Y., {Herbst}, E., {Roberts}, H., \& {Caselli}, P. 2005, \apj, 620,
  330

\bibitem[{Bacmann {et~al.}(2002)Bacmann, Lefloch, Ceccarelli, Castets,
  Steinacker, \& Loinard}]{Bacmann2002}
Bacmann, A., Lefloch, B., Ceccarelli, C., Castets, A., Steinacker, J., \&
  Loinard, L. 2002, \aap, 389, L6

\bibitem[{Bacmann {et~al.}(2003)Bacmann, Lefloch, Ceccarelli, Steinacker,
  Castets, \& Loinard}]{Bacmann2003}
Bacmann, A., Lefloch, B., Ceccarelli, C., Steinacker, J., Castets, A., \&
  Loinard, L. 2003, \apjl, 585, L55

\bibitem[{{Barsuhn}(1977)}]{Barsuhn1977}
{Barsuhn}, J. 1977, \aaps, 28, 453

\bibitem[{Bell {et~al.}(2006)Bell, Roueff, Viti, \& Williams}]{Bell2006a}
Bell, T.~A., Roueff, E., Viti, S., \& Williams, D.~A. 2006, \mnras, 371, 1865

\bibitem[{Bergin {et~al.}(2002)Bergin, Alves, Huard, \& Lada}]{Bergin2002}
Bergin, E.~A., Alves, J., Huard, T., \& Lada, C.~J. 2002, \apjl, 570, L101

\bibitem[{Boland \& de~Jong(1982)}]{Boland1982}
Boland, W., \& de~Jong, T. 1982, \apj, 261, 110

\bibitem[{Brown \& Millar(1989)}]{Brown1989a}
Brown, P.~D., \& Millar, T.~J. 1989, \mnras, 237, 661

\bibitem[{{Caselli} {et~al.}(1999){Caselli}, {Walmsley}, {Tafalla}, {Dore}, \&
  {Myers}}]{Caselli1999}
{Caselli}, P., {Walmsley}, C.~M., {Tafalla}, M., {Dore}, L., \& {Myers}, P.~C.
  1999, \apjl, 523, L165

\bibitem[{{Ceccarelli} {et~al.}(1998){Ceccarelli}, {Castets}, {Loinard},
  {Caux}, \& {Tielens}}]{Ceccarelli1998}
{Ceccarelli}, C., {Castets}, A., {Loinard}, L., {Caux}, E., \& {Tielens},
  A.~G.~G.~M. 1998, \aap, 338, L43

\bibitem[{{Cyburt} {et~al.}(2003){Cyburt}, {Fields}, \& {Olive}}]{Cyburt2003}
{Cyburt}, R.~H., {Fields}, B.~D., \& {Olive}, K.~A. 2003, Physics Letters B,
  567, 227

\bibitem[{Draine(1978)}]{Draine1978}
Draine, B.~T. 1978, \apjs, 36, 595

\bibitem[{Draine(2004)}]{Draine2004}
---. 2004, in Carnegie Observatories Centennial Symp.,
  Origin and Evolution of the Elements, ed. A.~McWilliam \& M.~Rauch
  (Cambridge: Cambridge Univ.~Press), 317

\bibitem[{Draine(2006)}]{Draine2006}
---. 2006, in ASP Conf.~Ser.~348, Astrophysics in the Far Ultraviolet: Five Years of Discovery with {\it FUSE}, ed. G.~Sonneborn, H.~W. Moos, \& B.-G. Andersson (San Francisco: ASP), 58

\bibitem[{Ellison {et~al.}(2007)Ellison, Prochaska, \& Lopez}]{Ellison2007}
Ellison, S.~L., Prochaska, J.~X., \& Lopez, S. 2007, \mnras, 380, 1245

\bibitem[{Federman \& Allen(1991)}]{Federman1991}
Federman, S.~R., \& Allen, M. 1991, \apj, 375, 157

\bibitem[{{Flower} {et~al.}(2004){Flower}, {Pineau des For{\^e}ts}, \&
  {Walmsley}}]{Flower2004}
{Flower}, D.~R., {Pineau des For{\^e}ts}, G., \& {Walmsley}, C.~M. 2004, \aap,
  427, 887

\bibitem[{{Flower} {et~al.}(2006{\natexlab{a}}){Flower}, {Pineau des
  For{\^e}ts}, \& {Walmsley}}]{Flower2006a}
---. 2006{\natexlab{a}}, \aap, 449, 621

\bibitem[{{Flower} {et~al.}(2006{\natexlab{b}}){Flower}, {Pineau des
  For{\^e}ts}, \& {Walmsley}}]{Flower2006b}
---. 2006{\natexlab{b}}, \aap, 456, 215

\bibitem[{{Gerin} {et~al.}(2001){Gerin}, {Pearson}, {Roueff}, {Falgarone}, \&
  {Phillips}}]{Gerin2001}
{Gerin}, M., {Pearson}, J.~C., {Roueff}, E., {Falgarone}, E., \& {Phillips},
  T.~G. 2001, \apjl, 551, L193

\bibitem[{{Hatchell}(2003)}]{Hatchell2003}
{Hatchell}, J. 2003, \aap, 403, L25

\bibitem[{{Hinze}(1975)}]{Hinze1975}
{Hinze}, J.~O. 1975, Turbulence (New York: McGraw-Hill)

\bibitem[{{Hirota} {et~al.}(2003){Hirota}, {Ikeda}, \& {Yamamoto}}]{Hirota2003}
{Hirota}, T., {Ikeda}, M., \& {Yamamoto}, S. 2003, \apj, 594, 859

\bibitem[{{Indriolo} {et~al.}(2007){Indriolo}, {Geballe}, {Oka}, \&
  {McCall}}]{Indriolo2007}
{Indriolo}, N., {Geballe}, T.~R., {Oka}, T., \& {McCall}, B.~J. 2007, \apj,
  671, 1736

\bibitem[{Jefferts {et~al.}(1973)Jefferts, Penzias, \& Wilson}]{Jefferts1973}
Jefferts, K.~B., Penzias, A.~A., \& Wilson, R.~W. 1973, \apjl, 179, L57

\bibitem[{{Kramer} {et~al.}(1999){Kramer}, {Alves}, {Lada}, {Lada}, {Sievers},
  {Ungerechts}, \& {Walmsley}}]{Kramer1999}
{Kramer}, C., {Alves}, J., {Lada}, C.~J., {Lada}, E.~A., {Sievers}, A.,
  {Ungerechts}, H., \& {Walmsley}, C.~M. 1999, \aap, 342, 257

\bibitem[{Lacour {et~al.}(2005)Lacour, Andr{\'e}, Sonnentrucker, Le~Petit,
  Welty, Desert, Ferlet, Roueff, \& York}]{Lacour2005}
Lacour, S., Andr{\'e}, M.~K., Sonnentrucker, P., Le~Petit, F., Welty, D.~E.,
  Desert, J.-M., Ferlet, R., Roueff, E., \& York, D.~G. 2005, \aap, 430, 967

\bibitem[{Le~Petit {et~al.}(2002)Le~Petit, Roueff, \& Le~Bourlot}]{LePetit2002}
Le~Petit, F., Roueff, E., \& Le~Bourlot, J. 2002, \aap, 390, 369

\bibitem[{Le~Teuff {et~al.}(2000)Le~Teuff, Millar, \& Markwick}]{LeTeuff2000}
Le~Teuff, Y.~H., Millar, T.~J., \& Markwick, A.~J. 2000, \aaps, 146, 157

\bibitem[{Lee {et~al.}(1996)Lee, Herbst, Pineau~des For{\^e}ts, Roueff, \&
  Le~Bourlot}]{Lee1996}
Lee, H.-H., Herbst, E., Pineau~des For{\^e}ts, G., Roueff, E., \& Le~Bourlot,
  J. 1996, \aap, 311, 690

\bibitem[{Lesaffre {et~al.}(2007)Lesaffre, Gerin, \& Hennebelle}]{Lesaffre2007}
Lesaffre, P., Gerin, M., \& Hennebelle, P. 2007, \aap, 469, 949

\bibitem[{Linsky {et~al.}(2006)Linsky, Draine, Moos, Jenkins, Wood, Oliveira,
  Blair, Friedman, Gry, Knauth, Kruk, Lacour, Lehner, Redfield, Shull,
  Sonneborn, \& Williger}]{Linsky2006}
Linsky, J.~L., Draine, B.~T., Moos, H.~W., Jenkins, E.~B., Wood, B.~E.,
  Oliveira, C., Blair, W.~P., Friedman, S.~D., Gry, C., Knauth, D., Kruk,
  J.~W., Lacour, S., Lehner, N., Redfield, S., Shull, J.~M., Sonneborn, G., \&
  Williger, G.~M. 2006, \apj, 647, 1106

\bibitem[{Lis {et~al.}(2002)Lis, Roueff, Gerin, Phillips, Coudert, van~der Tak,
  \& Schilke}]{Lis2002}
Lis, D.~C., Roueff, E., Gerin, M., Phillips, T.~G., Coudert, L.~H., van~der
  Tak, F.~F.~S., \& Schilke, P. 2002, \apjl, 571, L55

\bibitem[{Loinard {et~al.}(2001)Loinard, Castets, Ceccarelli, Caux, \&
  Tielens}]{Loinard2001}
Loinard, L., Castets, A., Ceccarelli, C., Caux, E., \& Tielens, A.~G.~G.~M.
  2001, \apjl, 552, L163

\bibitem[{{Loinard} {et~al.}(2002){Loinard}, {Castets}, {Ceccarelli},
  {Lefloch}, {Benayoun}, {Caux}, {Vastel}, {Dartois}, \&
  {Tielens}}]{Loinard2002}
{Loinard}, L., {Castets}, A., {Ceccarelli}, C., {Lefloch}, B., {Benayoun},
  J.-J., {Caux}, E., {Vastel}, C., {Dartois}, E., \& {Tielens}, A.~G.~G.~M.
  2002, \planss, 50, 1205

\bibitem[{{McCall} {et~al.}(2003){McCall}, {Huneycutt}, {Saykally}, {Geballe},
  {Djuric}, {Dunn}, {Semaniak}, {Novotny}, {Al-Khalili}, {Ehlerding},
  {Hellberg}, {Kalhori}, {Neau}, {Thomas}, {{\"O}sterdahl}, \&
  {Larsson}}]{McCall2003}
{McCall}, B.~J., {Huneycutt}, A.~J., {Saykally}, R.~J., {Geballe}, T.~R.,
  {Djuric}, N., {Dunn}, G.~H., {Semaniak}, J., {Novotny}, O., {Al-Khalili}, A.,
  {Ehlerding}, A., {Hellberg}, F., {Kalhori}, S., {Neau}, A., {Thomas}, R.,
  {{\"O}sterdahl}, F., \& {Larsson}, M. 2003, \nat, 422, 500

\bibitem[{Oliveira \& H{\'e}brard(2006)}]{Oliveira2006}
Oliveira, C.~M., \& H{\'e}brard, G. 2006, \apj, 653, 345

\bibitem[{{Parise} {et~al.}(2004){Parise}, {Castets}, {Herbst}, {Caux},
  {Ceccarelli}, {Mukhopadhyay}, \& {Tielens}}]{Parise2004}
{Parise}, B., {Castets}, A., {Herbst}, E., {Caux}, E., {Ceccarelli}, C.,
  {Mukhopadhyay}, I., \& {Tielens}, A.~G.~G.~M. 2004, \aap, 416, 159

\bibitem[{{Parise} {et~al.}(2005){Parise}, {Caux}, {Castets}, {Ceccarelli},
  {Loinard}, {Tielens}, {Bacmann}, {Cazaux}, {Comito}, {Helmich}, {Kahane},
  {Schilke}, {van Dishoeck}, {Wakelam}, \& {Walters}}]{Parise2005}
{Parise}, B., {Caux}, E., {Castets}, A., {Ceccarelli}, C., {Loinard}, L.,
  {Tielens}, A.~G.~G.~M., {Bacmann}, A., {Cazaux}, S., {Comito}, C., {Helmich},
  F., {Kahane}, C., {Schilke}, P., {van Dishoeck}, E., {Wakelam}, V., \&
  {Walters}, A. 2005, \aap, 431, 547

\bibitem[{{Parise} {et~al.}(2002){Parise}, {Ceccarelli}, {Tielens}, {Herbst},
  {Lefloch}, {Caux}, {Castets}, {Mukhopadhyay}, {Pagani}, \&
  {Loinard}}]{Parise2002}
{Parise}, B., {Ceccarelli}, C., {Tielens}, A.~G.~G.~M., {Herbst}, E.,
  {Lefloch}, B., {Caux}, E., {Castets}, A., {Mukhopadhyay}, I., {Pagani}, L.,
  \& {Loinard}, L. 2002, \aap, 393, L49

\bibitem[{Phillips \& Huggins(1981)}]{Phillips1981}
Phillips, T.~G., \& Huggins, P.~J. 1981, \apj, 251, 533

\bibitem[{Phillips \& Lis(2006)}]{Phillips2006}
Phillips, T.~G., \& Lis, D.~C. 2006, in ASP Conf.~Ser.~356,
  Revealing the Molecular Universe: One Antenna is Never Enough,
  ed. D.~C. Backer, J.~M. Moran, \& J.~L. Turner (San Francisco:
  ASP), 223

\bibitem[{Phillips \& Vastel(2003)}]{Phillips2003}
Phillips, T.~G., \& Vastel, C. 2003, in Chemistry as a Diagnostic of Star
  Formation, ed. C.~L. Curry \& M.~Fich (Ottawa: NRC Press), 3

\bibitem[{{Prandtl}(1925)}]{Prandtl1925}
{Prandtl}, L. 1925, Z.~Angew.~Math.~Mech., 5, 136

\bibitem[{{Prochaska} {et~al.}(2005){Prochaska}, {Tripp}, \&
  {Howk}}]{Prochaska2005}
{Prochaska}, J.~X., {Tripp}, T.~M., \& {Howk}, J.~C. 2005, \apjl, 620, L39

\bibitem[{{Prodanovi{\'c}} \& {Fields}(2008)}]{Prodanovic2008}
{Prodanovi{\'c}}, T., \& {Fields}, B.~D. 2008, Journal of Cosmology and
  Astroparticle Physics, 9, 3

\bibitem[{Rachford {et~al.}(2009)Rachford, Snow, Destree, Ross, Ferlet,
  Friedman, Gry, Jenkins, Morton, Savage, Shull, Sonnentrucker, Tumlinson,
  Vidal-Madjar, Welty, \& York}]{Rachford2009}
Rachford, B.~L., Snow, T.~P., Destree, J.~D., Ross, T.~L., Ferlet, R.,
  Friedman, S.~D., Gry, C., Jenkins, E.~B., Morton, D.~C., Savage, B.~D.,
  Shull, J.~M., Sonnentrucker, P., Tumlinson, J., Vidal-Madjar, A., Welty,
  D.~E., \& York, D.~G. 2009, \apjs, 180, 125

\bibitem[{Rachford {et~al.}(2002)Rachford, Snow, Tumlinson, Shull, Blair,
  Ferlet, Friedman, Gry, Jenkins, Morton, Savage, Sonnentrucker, Vidal-Madjar,
  Welty, \& York}]{Rachford2002}
Rachford, B.~L., Snow, T.~P., Tumlinson, J., Shull, J.~M., Blair, W.~P.,
  Ferlet, R., Friedman, S.~D., Gry, C., Jenkins, E.~B., Morton, D.~C., Savage,
  B.~D., Sonnentrucker, P., Vidal-Madjar, A., Welty, D.~E., \& York, D.~G.
  2002, \apj, 577, 221

\bibitem[{{Roberts} {et~al.}(2002){Roberts}, {Fuller}, {Millar}, {Hatchell}, \&
  {Buckle}}]{Roberts2002}
{Roberts}, H., {Fuller}, G.~A., {Millar}, T.~J., {Hatchell}, J., \& {Buckle},
  J.~V. 2002, \aap, 381, 1026

\bibitem[{Roberts {et~al.}(2003)Roberts, Herbst, \& Millar}]{Roberts2003}
Roberts, H., Herbst, E., \& Millar, T.~J. 2003, \apjl, 591, L41

\bibitem[{Roberts {et~al.}(2004)Roberts, Herbst, \& Millar}]{Roberts2004}
---. 2004, \aap, 424, 905

\bibitem[{Roberts \& Millar(2000{\natexlab{a}})}]{Roberts2000a}
Roberts, H., \& Millar, T.~J. 2000{\natexlab{a}}, \aap, 361, 388

\bibitem[{Roberts \& Millar(2000{\natexlab{b}})}]{Roberts2000b}
---. 2000{\natexlab{b}}, \aap, 364, 780

\bibitem[{{Roberts} \& {Millar}(2007)}]{Roberts2007}
---. 2007, \aap, 471, 849

\bibitem[{Rodgers \& Charnley(2001)}]{Rodgers2001}
Rodgers, S.~D., \& Charnley, S.~B. 2001, \apj, 553, 613

\bibitem[{{Roueff} {et~al.}(2007){Roueff}, {Herbst}, {Lis}, \&
  {Phillips}}]{Roueff2007b}
{Roueff}, E., {Herbst}, E., {Lis}, D.~C., \& {Phillips}, T.~G. 2007, \apjl,
  661, L159

\bibitem[{Roueff {et~al.}(2005)Roueff, Lis, van~der Tak, Gerin, \&
  Goldsmith}]{Roueff2005}
Roueff, E., Lis, D.~C., van~der Tak, F.~F.~S., Gerin, M., \& Goldsmith, P.~F.
  2005, \aap, 438, 585

\bibitem[{Roueff {et~al.}(2000)Roueff, Tin{\'e}, Coudert, Pineau~des
  For{\^e}ts, Falgarone, \& Gerin}]{Roueff2000}
Roueff, E., Tin{\'e}, S., Coudert, L.~H., Pineau~des For{\^e}ts, G., Falgarone,
  E., \& Gerin, M. 2000, \aap, 354, L63

\bibitem[{Snow {et~al.}(2008)Snow, Ross, Destree, Drosback, Jensen, Rachford,
  Sonnentrucker, \& Ferlet}]{Snow2008}
Snow, T.~P., Ross, T.~L., Destree, J.~D., Drosback, M.~M., Jensen, A.~G.,
  Rachford, B.~L., Sonnentrucker, P., \& Ferlet, R. 2008, \apj, 688, 1124

\bibitem[{Solomon \& Woolf(1973)}]{Solomon1973}
Solomon, P.~M., \& Woolf, N.~J. 1973, \apjl, 180, L89

\bibitem[{{Steigman} {et~al.}(2007){Steigman}, {Romano}, \&
  {Tosi}}]{Steigman2007}
{Steigman}, G., {Romano}, D., \& {Tosi}, M. 2007, \mnras, 378, 576

\bibitem[{{Tafalla} {et~al.}(2004){Tafalla}, {Myers}, {Caselli}, \&
  {Walmsley}}]{Tafalla2004b}
{Tafalla}, M., {Myers}, P.~C., {Caselli}, P., \& {Walmsley}, C.~M. 2004, \aap,
  416, 191

\bibitem[{{Taylor}(1915)}]{Taylor1915}
{Taylor}, G.~I. 1915, Royal Society of London Philosophical Transactions Series
  A, 215, 1

\bibitem[{{Tielens}(1983)}]{Tielens1983}
{Tielens}, A.~G.~G.~M. 1983, \aap, 119, 177

\bibitem[{{Tin{\'e}} {et~al.}(2000){Tin{\'e}}, {Roueff}, {Falgarone}, {Gerin},
  \& {Pineau des For{\^e}ts}}]{Tine2000}
{Tin{\'e}}, S., {Roueff}, E., {Falgarone}, E., {Gerin}, M., \& {Pineau des
  For{\^e}ts}, G. 2000, \aap, 356, 1039

\bibitem[{{Tosi}(1996)}]{Tosi1996}
{Tosi}, M. 1996, in ASP Conf.~Ser.~98, From Stars to Galaxies: the Impact
  of Stellar Physics on Galaxy Evolution, ed. C. {Leitherer},
  U. {Fritze-von-Alvensleben}, \& J. {Huchra} (San Francisco: ASP), 299

\bibitem[{{Turner}(2001)}]{Turner2001}
{Turner}, B.~E. 2001, \apjs, 136, 579

\bibitem[{van~der Tak {et~al.}(2002)van~der Tak, Schilke, M{\"u}ller, Lis,
  Phillips, Gerin, \& Roueff}]{vanderTak2002}
van~der Tak, F.~F.~S., Schilke, P., M{\"u}ller, H.~S.~P., Lis, D.~C., Phillips,
  T.~G., Gerin, M., \& Roueff, E. 2002, \aap, 388, L53

\bibitem[{{Vastel} {et~al.}(2003){Vastel}, {Phillips}, {Ceccarelli}, \&
  {Pearson}}]{Vastel2003}
{Vastel}, C., {Phillips}, T.~G., {Ceccarelli}, C., \& {Pearson}, J. 2003,
  \apjl, 593, L97

\bibitem[{Vastel {et~al.}(2004)Vastel, Phillips, \& Yoshida}]{Vastel2004}
Vastel, C., Phillips, T.~G., \& Yoshida, H. 2004, \apjl, 606, L127

\bibitem[{{Vasyunin} {et~al.}(2009){Vasyunin}, {Semenov}, {Wiebe}, \&
  {Henning}}]{Vasyunin2009}
{Vasyunin}, A.~I., {Semenov}, D.~A., {Wiebe}, D.~S., \& {Henning}, T. 2009,
  \apj, 691, 1459

\bibitem[{{Watson}(1976)}]{Watson1976}
{Watson}, W.~D. 1976, Reviews of Modern Physics, 48, 513

\bibitem[{Willacy(2007)}]{Willacy2007}
Willacy, K. 2007, \apj, 660, 441

\bibitem[{Willacy {et~al.}(2002)Willacy, Langer, \& Allen}]{Willacy2002}
Willacy, K., Langer, W.~D., \& Allen, M. 2002, \apjl, 573, L119

\bibitem[{{Willacy} {et~al.}(1998){Willacy}, {Langer}, \&
  {Velusamy}}]{Willacy1998}
{Willacy}, K., {Langer}, W.~D., \& {Velusamy}, T. 1998, \apjl, 507, L171

\bibitem[{{Williams} {et~al.}(1998){Williams}, {Bergin}, {Caselli}, {Myers}, \&
  {Plume}}]{Williams1998}
{Williams}, J.~P., {Bergin}, E.~A., {Caselli}, P., {Myers}, P.~C., \& {Plume},
  R. 1998, \apj, 503, 689

\bibitem[{{Wood} {et~al.}(2004){Wood}, {Linsky}, {H{\'e}brard}, {Williger},
  {Moos}, \& {Blair}}]{Wood2004}
{Wood}, B.~E., {Linsky}, J.~L., {H{\'e}brard}, G., {Williger}, G.~M., {Moos},
  H.~W., \& {Blair}, W.~P. 2004, \apj, 609, 838

\bibitem[{Xie {et~al.}(1995)Xie, Allen, \& Langer}]{Xie1995}
Xie, T., Allen, M., \& Langer, W.~D. 1995, \apj, 440, 674

\end{thebibliography}
\end{document}